\documentclass{article}
\usepackage{graphicx} 
\usepackage{amsmath}
\usepackage[a4paper, margin=1in]{geometry}
\usepackage{jheppub}

\usepackage{lscape}
\usepackage{tikz}
\usepackage{array}
\usepackage{hyperref}
\usepackage{amsfonts}
\usepackage{amssymb}
\usepackage{amsmath, amsthm}
\usepackage{bbm} 
\usepackage{graphicx}
\usepackage{caption}
\usepackage{mathtools}
\usepackage{imakeidx}
\usepackage[utf8]{inputenc}
\usepackage[T1]{fontenc}
\usepackage{booktabs}
\usepackage{multirow}
\usepackage{enumitem}

\newcommand{\D}{{\mathcal{D}}}

\newcommand{\appref}[1]{\hyperref[#1]{Appendix~\ref*{#1}}}

\title{Projecting Gravitational Fluctuations onto Near-Horizon Throats}
\author[1,2]{Alejandro Cabo-Bizet,}
\author[3]{Vasil Dimitrov,}
\author[1,2]{Dario Martelli}
\author[1,2]{and Lorenzo Ruggeri}
\affiliation[1]{Dipartimento di Matematica ``Giuseppe Peano'', Università di Torino, Via Carlo Alberto 10, 10123 Torino, Italy}
\affiliation[2]{INFN, Sezione di Torino, Via Pietro Giuria 1, 10125 Torino, Italy}
\affiliation[3]{INRNE, Bulgarian Academy of Sciences, Tsarigradsko Chaussee 72, 1784 Sofia, Bulgaria}
\emailAdd{acbizet@gmail.com}
\emailAdd{vasild@inrne.bas.bg}
\emailAdd{dario.martelli@unito.it}
\emailAdd{lorenzo.ruggeri@unito.it}

\abstract{We show that, for stationary geometries whose linearized fluctuation equations separate into radial and angular Heun equations, a pair of diffeomorphisms projects the full problem onto two coupled, isospectral Gauss hypergeometric problems in complementary spacetime regions. This projection requires Robin boundary conditions to be imposed at the intersection between the complementary regions, which are uniquely fixed by requiring the two diffeomorphisms to match there. The method provides an independent derivation of recent results obtained from the correspondence between Heun connection coefficients and instanton partition functions in the Nekrasov--Shatashvili limit. As an application, we identify a subset of modes in Kerr–de Sitter and Kerr–anti-de Sitter black holes that, in the near-extremal limit, continuously reduce to a basis of the Schwarzian/Jackiw--Teitelboim functional space of gravitational fluctuations.}

\begin{document}

\maketitle

\section{Introduction}
Many extremal black holes admit near-horizon decoupling limits in which the geometry exhibits an enhanced symmetry. As extremality is approached, the corresponding near-horizon region develops a parametrically long throat, often containing a locally AdS$_2$ sector.\footnote{In the strict extremal limit, the resulting near-horizon geometry is typically a possibly warped product or fibration of AdS$_2$ over a $(D-2)$-dimensional space.} This feature has been thoroughly studied in the context of string theory, holography and quantum gravity~\cite{Strominger:1998yg,MaldacenaMichelsonStrominger1999,BardeenHorowitz1999,Sen:2005wa,Sen:2007qy,Kunduri:2013gce}. In these cases, the corresponding low-energy dynamics is simpler than that of the full geometry, owing to the emergence of an $\mathrm{SL}(2,\mathbb{R})$ symmetry---the isometry of the AdS$_2$ factor. Long near-horizon throats have been established as a universal organizing framework for the study of quantum aspects of near-extremal black holes \cite{Sen:2008yk,Sen:2008vm,Almheiri:2014cka,Jensen:2016pah,Maldacena:2016upp,Engelsoy:2016xyb,Iliesiu:2020qvm,Iliesiu:2022onk}. A central open question is whether the asymptotic long throat perspective, supplemented by suitable boundary data inherited from the exterior region complementary to the long-throat \cite{Faulkner:2009wj,Ren:2012hg}, can reproduce arbitrarily high orders in the small-temperature expansion of the full fluctuation spectrum.

This work answers this question for a broad class of geometries, including four- and five-dimensional black holes and black branes. We show that all perturbative corrections in the small-temperature expansion of the linear fluctuation spectrum can be computed entirely within the near-horizon throat, provided that the effect of the exterior region is retained through appropriate boundary conditions at the edge of the throat.\footnote{Equivalently, the same corrections may be computed in the outer region. In that description, boundary conditions imposed in the overlap region encode the information carried by the near-horizon region.} We also provide a procedure for reconstructing the complete wave-functions from their near-horizon profiles, order by order in the small-temperature expansion. Thus, the full fluctuation problem admits an all-orders perturbative reduction to two complementary near-horizon and outer descriptions, coupled through their boundary data. 

The mechanism underlying this reduction, which in a sense~\cite{Maldacena:1997re} is holographic, is most transparent at the level of the separated fluctuation equations. For many four- and five-dimensional black holes and black branes, the radial and angular equations are second-order ordinary differential equations of Heun type, or arise as confluent limits thereof. Their spectra and mode functions are determined by connection problems between local solutions at two physically distinguished singular points, where the appropriate boundary conditions are imposed.\footnote{For a Kerr--de Sitter black hole, for example, the relevant singular points of the radial equation correspond to the event and cosmological horizons.} In holographic applications, the corresponding connection and spectral data determine the analytic structure of boundary thermal correlators and can encode information about the black hole interior and singularity \cite{Fidkowski:2003nf,Festuccia:2005pi,Dodelson:2023vrw,Dodelson:2025jff}.

These connection problems have recently revealed a rich algebraic structure and unexpected relations to two-dimensional conformal field theory, supersymmetric gauge theory, and integrable systems. The relevance of two-dimensional conformal symmetry to Kerr black holes was first emphasized in the Kerr/CFT correspondence and in the subsequent discovery of a hidden conformal symmetry of the low-frequency wave equation \cite{Guica:2008mu,Castro:2010fd}. The separated Kerr equations were later related more directly to isomonodromic deformation problems: their scattering and spectral data can be expressed in terms of Painlev\'e tau functions and, through the conformal-block expansion of these functions, in terms of Liouville CFT data \cite{Castro:2013lba,Novaes:2014lha,CarneirodaCunha:2015hzd,CarneirodaCunha:2015qln}. The mathematical foundation of this correspondence is closely related to the connection among isomonodromic tau functions, Liouville conformal blocks, and Nekrasov instanton partition functions \cite{Gamayun:2012ma,Gamayun:2013auu,Litvinov:2013sxa,Iorgov:2014vla}.

More recently, these relations have led to exact and perturbative methods for computing black hole spectra and Heun connection coefficients using Seiberg--Witten theory and instanton partition functions in the Nekrasov--Shatashvili limit \cite{Aminov:2020yma,Bonelli:2021uvf,Bonelli:2022ten,Lisovyy:2022flm}. Complementary approaches have exploited the associated integrable structures, including thermodynamic Bethe ansatz methods \cite{fioravanti2021newmethodexactresults}. A variety of extensions and applications have appeared in recent years \cite{Kapec:2023ruw,Kapec:2024zdj,Rakic:2023vhv,Kolanowski:2024,Mariani:2025hee,Castro:2025itb,PandoZayas:2026vbg, Arnaudo:2025uos,Arnaudo:2025kit,Arnaudo:2026der,Aniceto:2026zmc}.

In these approaches, once a local solution is selected by the physical boundary condition at one endpoint of the relevant domain, imposing the boundary condition at the other endpoint reduces the spectral problem to a single quantization condition, which can often be written in Bethe-ansatz form. From the perspective of the Bethe/gauge correspondence, this condition is encoded in the Nekrasov--Shatashvili limit of the partition function of a four-dimensional $\mathcal{N}=2$ supersymmetric gauge theory. For Heun and confluent-Heun problems, the relevant theories are typically rank-one $SU(2)$ theories with different numbers of fundamental hypermultiplets $N_f$. Their rank-one nature implies the existence of a single independent quantum period and, correspondingly, a single independent Bethe-type quantization condition for each separated spectral problem.

Motivated by the questions raised above, we follow  an approach that is complementary to those discussed earlier. We exploit the fact that, under suitable changes of coordinates, a Heun equation can be reduced to Gauss hypergeometric form in an appropriate degeneration limit.\footnote{A related degeneration picture appears in \cite{Nekrasov:2011bc}, where opers on a nearly degenerate punctured Riemann surface are approximated by hypergeometric opers on three-punctured components, glued through plumbing parameters, with corrections treated perturbatively.} At the level of the differential operator, this suggests a possible isospectral relation between the Heun and hypergeometric problems. The spectrum, however, depends crucially on the boundary conditions. Since the degeneration changes the asymptotic behaviour at one of the endpoints, a prescribed boundary condition for the Heun problem is not automatically preserved. It is therefore not evident whether the physical Heun spectrum can be continuously related to that of an auxiliary Gauss hypergeometric problem.

Recent extensions of the Gelfand--Yaglom theorem and of the Denef--Hartnoll--Sachdev formula \cite{Denef:2009kn} provide encouraging evidence in this direction. In particular, \cite{Arnaudo:2024rhv,Arnaudo:2025btb} showed that one-loop determinants of Heun-type kinetic operators admit smooth degenerations to determinants of Gauss hypergeometric operators. These results, however, do not determine the boundary conditions inherited by the limiting hypergeometric problem, nor do they establish whether arbitrarily high-order corrections away from the hypergeometric limit can be reconstructed entirely within the limiting description.

We settle these questions by constructing two complementary hypergeometric descriptions of the full Heun problem. For simplicity, in this paper we focus on the equations governing linear fluctuations of Kerr--de Sitter and Kerr--anti-de Sitter black holes \cite{Teukolsky:1972my,Batic:2007it}, although the method applies more generally. In the zero-temperature degeneration limit, the full radial problem splits into inner and outer hypergeometric patches. The information carried by the region omitted from either patch is not lost; instead, it is encoded in twisted, or Robin-type, boundary conditions imposed in the matching region. These boundary conditions generically break the $\mathrm{SL}(2,\mathbb{R})$ symmetry of the auxiliary hypergeometric problem.

We then show that suitable diffeomorphisms remove the interaction terms from the bulk differential operators order by order in the small-temperature expansion. Their effect is transferred to the matching data between the two hypergeometric patches, and may equivalently be viewed as a defect localized in the overlap region. This construction determines arbitrarily high-order corrections to connection coefficients and, consequently, to the fluctuation spectrum and wave-functions. The resulting asymptotic expansions may contain non-analytic contributions, including fractional powers and logarithms of the small parameter. As a consistency check, we reproduce the corrections predicted by Nekrasov instanton counting in the Nekrasov--Shatashvili limit. 

In the holographic setting, our results address recent questions regarding Schwarzian near-zero modes \cite{Moitra:2019bub,Arnaudo:2024bbd,Arnaudo:2025btb,Kapec:2023ruw,Rakic:2023vhv,Kapec:2024zdj,Kolanowski:2024,Mariani:2025hee,Bac:2026eqj,Cheng:2026ull,Despontin:2026xzg}. As an application, we identify, in Kerr backgrounds, a subspace of gravitational fluctuations whose leading near-extremal limit coincides with the Schwarzian/Jackiw--Teitelboim near-zero-mode sector. We carry out the analysis directly in Lorentzian signature, using a notion of pseudo-square-integrability that follows from a non-Hermitian bilinear pairing defined patchwise over the outer and inner regions.

Beyond the holographic applications that constitute the main focus of this paper, the same method can be used to compute post-Minkowskian corrections to gravitational interactions and post-Newtonian corrections to gravitational waveforms \cite{Bautista:2023sdf,Cipriani:2025ikx,CaronHuot:2025BornSeriesTidal,CombaluzierSzteinsznaider:2026DynamicalTidal}. Existing approaches typically require the evaluation of integrals, which may be organized diagrammatically through a Born series. By contrast, the method presented here bypasses the need to compute such integrals.

\paragraph{Outline of the Paper.}
The rest of the paper is organized as follows. In the remaining part of the introduction, \autoref{sec:Strategy}, we summarize the main results and introduce the functional space of fluctuations. We also explain the role of the inner and outer hypergeometric descriptions and of the corresponding Robin boundary conditions. In \autoref{sec:2}, we study the leading large-$t$ limit of the Heun problem. We show that it splits into two complementary hypergeometric problems, defined in the outer and inner regions, and discuss how their discrete bases are glued in the matching region. In \autoref{sec:3}, we introduce diffeomorphisms which remove the interaction terms order by order in the large-$t$ expansion. This gives two isospectral hypergeometric problems whose boundary data encode the effect of the full Heun geometry. In \autoref{sec.4}, we apply the construction to gravitational fluctuations of Kerr black holes. We identify the low-energy square-integrable modes which reduce, in the cold limit, to the Schwarzian/Jackiw--Teitelboim zero-mode sector. Finally, in \autoref{sec.5}, we summarize the results and discuss possible extensions. Technical details and complementary computations are collected in the appendices.

\subsection{Summary of Results}\label{sec:Strategy}
We end this introduction by presenting a brief overview of the method used throughout the paper, and the main results. Our goal is to find the spectrum and wave-functions of differential equations of the form
\begin{equation}\label{eq:OpOriginal}
    \mathcal{L}_2=\left(\frac{\dd^2}{\dd z^2}+Q(z)\right)\,,
\end{equation}
where the potential $Q(z)$ contains singularities in the complex $z$-plane. The explicit case we will consider is that of a Heun equation, with four regular singularities at $z=0,1,t,\infty$, but we will also comment on equations with more than four singularities and on equations with irregular singularities. Any second-order differential equation admits two independent local Frobenius solutions around each of its regular singular points. The space of functions we will consider satisfies, at the endpoints $z=0$ and $z=\infty$, Frobenius boundary conditions, meaning that the wave-function selects only one, preferably the subleading, out of the two local solutions.

For path integral computations, we introduce a functional space that we will call the \emph{Gravitational functional space of linear fluctuations}\footnote{In physically relevant examples where a continuum exists above the discrete spectrum—including the de Sitter black hole solutions considered here—we restrict our analysis to normalisable discrete states and transitions entirely within the discrete sector. We assume that the relevant interactions do not mix the discrete and continuous spectra, so that matrix elements connecting the two sectors vanish. Consequently, within this assumption, neither real transitions into scattering continuum states nor virtual corrections mediated by intermediate continuum modes will be considered. Consequently, in contradistinction with related discussions, as for instance~\cite{Lestingi:2026peq} which works over \emph{larger} functional spaces, all resolutions of the identity used in this paper will be over the functional space $\mathcal{H}$, which by definition, does not include states spanned by eigenfunctions with continuous eigenvalues.   }
\begin{equation}\label{eq:FunctionalSpace}
    \mathcal{H}(t) =\bigg\{|\Phi\rangle:\quad|\Phi\rangle=\sum_n  c_n(t)|\Phi_n\rangle\bigg\}\,.
\end{equation}
This is a space over which a path integral of linear fluctuations around the gravitational background can be computed. The \emph{twisted basis elements} $|\Phi_n\rangle$ form a basis in terms of which any solution $|\Phi\rangle$ of \eqref{eq:OpOriginal} satisfying Frobenius boundary conditions at both endpoints, $z=0$ and $z=\infty$, can be expanded. Further details are provided below. 

\paragraph{Hypergeometric Functions and Boundary Conditions in the Inner/Outer Regions.}
We consider Schrödinger potentials of the form
\begin{equation}
    Q(z)=Q_0(z)+Q_\text{int}(z,t)\,,
\end{equation}
and take $1/t$ to be a deformation parameter in the large-$t$ limit. The unperturbed potential $Q_0(z)$ is chosen to be a hypergeometric potential with three regular singular points. By contrast, $Q_\text{int}(z,t)$ is treated as a deformation which may introduce additional regular singularities, as in an ordinary Heun equation, or irregular singularities, as in a confluent-Heun problem.

The explicit expression of $Q_0$ depends on whether $z\ll t$, which we dub \emph{outer region}, or $z\gg t$, which we dub \emph{inner region}. Starting from the \emph{outer region}, one finds\footnote{Eventually, we will assume the coefficients $a_0$, $a_1$ and $\lambda^\text{out}$ to depend implicitly on $t$ with regular values at $t=\infty$.}
\begin{equation}
    Q_0^\text{out}(z)=Q_0^\text{out}(z,\lambda^\text{out})=\frac{\frac{1}{4}-a_0^2}{z^2}+\frac{\frac{1}{4}-a_1^2}{(z-1)^2}-\frac{\lambda^\text{out}}{z(z-1)}\,.
\end{equation}
The interaction potential can be written as a series in $1/t$ 
\begin{equation}
    Q_\text{int}^\text{out}(z,t)=\sum_{n=0}^{\infty} \frac{a_n^\text{out}(z)}{t^{n+1}}\,.
\end{equation}
Thus, we assume that in the outer region, the interaction potential is a small deformation of a hypergeometric equation.

The outer hypergeometric operator at $t=\infty$, and fixed $z$, is given by
\begin{equation}\label{eq:OuterProjection} 
    \mathcal{L}_2^{\text{out},(0)}(z)=\frac{\dd^2}{\dd z^2}+Q_0^\text{out}(z)\,.
\end{equation}
Note that it does not inherit the same asymptotic behaviour at $z=\infty$ as the original operator $\mathcal{L}_2$ at $z=\infty$. The reason is that the asymptotic form of the original potential $Q$ changes in the $t\to\infty$ expansion at fixed $z$.\footnote{Also, we are restricting to $z\ll t$, which means that we are excluding the portion of the original range of $z$ containing the original singularity at $z=\infty$.} That is, the limits $t\to\infty$ and $z\to\infty$ do not commute. This non-commutativity leads to two complementary hypergeometric descriptions. 

The second hypergeometric description arises in the \emph{inner region}, where we take $z$ to scale with $t$. A natural coordinate to describe this inner region is $\zeta=z/t$. Expanding at $t\to \infty$, at fixed $\zeta$, we obtain what we call the inner hypergeometric operator
\begin{equation}
    \mathcal{L}_2^{\text{in},(0)}(\zeta):=\frac{\dd^2}{\dd\zeta^2}+Q_0^\text{in}(\zeta)\,.
\end{equation}
Note that this differential operator cannot be obtained from $\mathcal{L}_2^{\text{out},(0)}(z)$ \eqref{eq:OuterProjection} by substituting $z=\zeta\,t$.

The basis elements $|\Phi_n\rangle$ are constructed patchwise, computing their outer and inner profiles and then functionally matching them at the intersection region. They are Gauss-hypergeometric functions in both outer and inner regions but, unlike the cases usually studied, they do not form $\mathrm{SL}(2,\mathbb{R})$ representations. To compute the spectrum of eigenvalues of $\Phi_n$, and the $c_n(t)$, in terms of their a priori prescribed leading asymptotic behaviour $c_n(t\sim\infty)$, one needs to go higher in the large-$t$ expansion.

In the two large-$t$ expansions, at fixed $z$ or $\zeta$, the boundary conditions are modified as follows:
\begin{equation*}\begin{gathered}
    \mathcal{L}_2:\;\text{Frobenius b.c. at both } z=0,\infty\\[0.5em]
    \Downarrow\\[0.5em]
    \mathcal{L}_2^{\text{out},(0)}:\;\left\{\begin{array}{c}
    \text{Robin b.c. at } z=\infty  \\
    \text{Frobenius b.c. at } z=0
    \end{array}\right.\quad\text{ and }\quad \mathcal{L}_2^{\text{in},(0)}:\;\left\{\begin{array}{c}
    \text{Frobenius b.c. at } \zeta=\infty  \\
    \text{Robin b.c. at } \zeta=0
    \end{array}\right.\,.
\end{gathered}\end{equation*}
By Robin boundary conditions we mean that, instead of demanding that one of the two local Frobenius solutions vanishes, a linear combination of the two solutions vanishes. Note that, even after imposing the correct Robin-Frobenius boundary conditions, the leading hypergeometric description still requires dealing with the computation of further $1/t$ corrections via a Born-series expansion of the radiative corrections, organized in terms of Feynman diagrams.

\paragraph{The Isospectral Projections to Outer/Inner Regions.}
We find that all such radiative corrections can be isospectrally disconnected by working with complex space-time coordinates $Z$ and $\wp$ for the outer and inner regions, respectively. The relation between the earlier coordinates reads
\begin{equation}
    z=F^\text{out}[Z]=Z+\mathcal{O}\left(\frac{1}{t}\right), \qquad \zeta=F^\text{in}[\wp]=\wp+\mathcal{O}\left(\frac{1}{t}\right)\,,
\end{equation}
at leading order in the large-$t$ expansion. Higher-order terms will be determined in \autoref{sec:3}. Importantly, these diffeomorphisms must match
\begin{equation}
    F^\text{out}[Z]=t\,F^\text{in}[\wp]
\end{equation}
in the asymptotic region where the outer and inner patches overlap
\begin{equation}\label{eq:MatchingRegionIntro}
    Z\gg1, \quad \wp\ll 1,\quad  \frac{Z}{t}\ll1, \quad \wp\,t \gg 1\,.
\end{equation}
We call the overlap between the two patches the \emph{matching region}.

Consider the action of $\mathcal{L}_2$ \eqref{eq:OpOriginal} on a wave-function
\begin{equation}
    \left[\frac{\dd^2}{\dd z^2}+Q\left(z\right)\right]\Psi(z)=0\,.
\end{equation}
Let us focus on the outer region. A generic change of the coordinate $z=F(Z)$ does not preserve the Schrödinger form unless the wave-function is also rescaled. The correct transformation is
\begin{equation}\label{eq:InducedActionwave-function}
    \Psi^\text{out}(F^\text{out}(Z))=\sqrt{(F^\text{out})'(Z)}\,\psi^\text{out}(Z).
\end{equation}
Then the transformed equation is again in Schrödinger form
\begin{equation}
    \left[\frac{\dd^2}{\dd Z^2}+\widetilde{Q}^\text{out}\left(Z\right)\right]\psi^\text{out}(Z)=0\,,
\end{equation}
where
\begin{equation}\label{eq:TransformationPotential}
    \widetilde{Q}^\text{out}\left(Z\right)=((F^\text{out})')^2Q\left(F^\text{out}\right)+\frac{1}{2}\{F^\text{out},Z\}
\end{equation}
and
\begin{equation}
    \{F^\text{out},Z\}=\frac{(F^\text{out})'''}{(F^\text{out})'}-\frac{3}{2}\left(\frac{(F^\text{out})''}{(F^\text{out})'}\right)^2
\end{equation}
is the Schwarzian derivative.

The diffeomorphism $F^\text{out}$ projects the Heun problem to a hypergeometric problem if
\begin{equation}
    \widetilde{Q}^\text{out}=Q_0^\text{out}\,.
\end{equation}
At the technical level, our goal is therefore to solve the following equation
\begin{equation}\label{eq:TransformationPotentialOuter}
    Q_0^\text{out}\left(Z\right)=((F^\text{out})')^2Q(z)+\frac{1}{2}\{F^\text{out},Z\}\,,
\end{equation}
in the large-$t$ expansion. In the inner region the discussion is analogous and we find the following equation to solve
\begin{equation}\label{eq:TransformationPotentialInner}
    Q_{0}^\text{in}\left(\wp\right)={\bigl(({F^\text{in}})^{'}\bigr)}^2 Q\left(z\right)+\frac{1}{2}\{F^\text{in},\wp\}\,.
\end{equation}
In this paper we focus on analytic procedures, but this can also be implemented numerically. We leave that to future work. 

An important technical observation of our paper is that the spectrum and wave-functions of $\mathcal{L}_2$, defined by Frobenius-Frobenius boundary conditions at $z=0$ and $z=\infty$, project to a \emph{synchronized} system of two dual hypergeometric problems with Robin-Frobenius boundary conditions. This projection is implemented by the diffeomorphisms $F^\text{out}$ and $F^\text{in}$, respectively. The large-$t$ expansion of the Heun spectrum remains invariant upon such projection because the diffeomorphisms are regular at all orders in the large-$t$ expansion. The relation between the wave-functions is non-trivial, though, as it is given by the large-$t$ expansions of \eqref{eq:TransformationPotentialOuter} and \eqref{eq:TransformationPotentialInner}.

\paragraph{Localization of the Wave-Function to Outer/Inner Regions.}
We also define the basis elements $|\Phi_n\rangle$ defining the functional space \eqref{eq:FunctionalSpace}. We do so patchwise 
\begin{equation}\label{eq:patchwiseDefinitonBasis}
    |\Phi_n\rangle=\left\{\begin{array}{cc}
        |\phi_n^\text{out}\rangle & \qquad\text{outer region} \\
        |\phi_n^\text{in}\rangle & \qquad\text{inner region}
    \end{array}\right.\,,
\end{equation}
in terms of two basis $\{|\phi_n^\text{out}\rangle\}$ and $\{|\phi_n^\text{in}\rangle\}$ living in the complementary regions of spacetime. By construction, these basis elements are normalisable with respect to a non-Hermitian bilinear pairing, defined patchwise by integration along a complex path in the radial variable. To distinguish this notion from ordinary square-integrability on the real radial variable, we refer to $\mathcal{H}$ in \eqref{eq:FunctionalSpace} as the functional space of \emph{pseudo-square-integrable fluctuations}. It is spanned by the pseudo-square-integrable states $\lvert\Phi_n\rangle$ and their linear combinations. Further details will be provided below.

In the overlap region \eqref{eq:MatchingRegionIntro}, by construction, these two local basis elements have the same asymptotics, with appropriate choices of normalization constants
\begin{equation}
    \langle Z|\phi_n^\text{out}\rangle=\langle\wp|\phi_n^\text{in}\rangle\,.
\end{equation}
We will find that, at a generic order in the large-$t$ expansion, neither the outer nor the inner projections of the twisted basis elements form representations of $\mathrm{SL}(2,\mathbb{R})$. Instead, $\mathrm{SL}(2,\mathbb{R})$ representations are defined when one imposes Frobenius-Frobenius boundary conditions in the hypergeometric projections.

Only specific subsets of basis elements $|\Phi_n\rangle$, when projected to one of the hypergeometric patches, may furnish a representation of $\mathrm{SL}(2,\mathbb{R})$ at leading order in the large-$t$ expansion. In those cases the Robin asymptotic behaviour of the corresponding basis element asymptotes to the correct square integrable Frobenius behaviour in one of the two patches, at leading order in the large-$t$ expansion. However, this cannot happen in both the outer and inner regions simultaneously. For the other patch, the boundary conditions remain Frobenius-Robin at leading order in the large-$t$ expansion and thus, $\mathrm{SL}(2,\mathbb{R})$ remains broken there. When this happens we say that such a subset of modes \emph{localizes} to the patch where the $\mathrm{SL}(2,\mathbb{R})$ symmetry remains unbroken at leading order in the large-$t$ expansion. 

In the last section we will use our construction to identify low-energy excitations that localize to the inner region of Kerr black holes, for both asymptotically dS and AdS geometries. We will do so at leading order in the $1/t$ expansion, which corresponds to the first small-temperature corrections. We will explicitly show that the corresponding leading asymptotic behaviour of such modes is that of the tensor and vector zero modes in AdS$_2$ associated with the Schwarzian. The very same gravitational modes were originally discussed in \cite{Iliesiu:2022onk} within the context of asymptotically flat four-dimensional Reissner-Nordstrom black holes. Here, we demonstrate their emergence, analytically, from the perspective of the complete Kerr geometry and working in Lorentzian signature.

\section{Leading-Order Analysis}\label{sec:2}
In this section we start by introducing the differential equations that play a key role in the rest of this work. As already discussed in the introduction, we consider differential equations which admit an expansion in a small parameter and such that, at leading order, they reduce to a hypergeometric equation. We also compute, at leading order, the spectrum and define the twisted basis elements for any solution. We perform this analysis in two complementary regions, admitting two inequivalent expansions of the differential equation. On the overlap between the two regions, we impose non-Frobenius boundary conditions and require the two solutions to patch smoothly. 

Let us start by defining the class of differential operators that will be studied in detail in the rest of this paper. While the procedure outlined in this work applies whenever a differential operator reduces to a hypergeometric equation by taking some parameters to be small, we will focus on the simplest instance of a Heun differential equation 
\begin{equation}\label{eq:Heun}
    \frac{\dd^2\Psi}{\dd z^2}+\left[\frac{\frac{1}{4}-a_0^2}{z^2}+\frac{\frac{1}{4}-a_1^2}{(z-1)^2}+\frac{\frac{1}{4}-a_t^2}{(z-t)^2}-\frac{\frac{1}{2}-a_1^2-a_t^2-a_0^2+a_\infty^2+u}{z(z-1)}+\frac{u}{z(z-t)}\right]\Psi(z)=0\,.
\end{equation}
This equation has four regular singularities at $z=0,1,t,\infty$. 

By taking the large-$t$ limit, the Heun equation reduces to a hypergeometric equation plus an interaction term. Denoting the Heun differential operator as 
\begin{equation}\label{eq.L2}
    \mathcal{L}_2=\frac{\dd^2}{\dd z^2}+Q(z)\,,
\end{equation}
the potential splits as $Q=Q_0+Q_\text{int}$, where the interaction term is suppressed by a factor of $1/t$. We can then introduce, at leading order, the hypergeometric operator
\begin{equation}\label{eq.hypersec2}
    \mathcal{L}_2^{(0)}=\frac{\dd^2}{\dd z^2}+Q_0(z)\,,
\end{equation}
with three regular singularities.

The hypergeometric part defines the auxiliary spectral problem, while the interaction encodes corrections in $1/t$ to the full operator. The explicit expression of the expansion depends on whether we take $z\ll t$ or $z\gg t$. The two expansions are valid in complementary regions, which together provide a natural decomposition of the $z$-plane into an \emph{outer} and an \emph{inner} region.

\subsection{The Outer Hypergeometric Projection}
Let us start by considering the outer region for $z\ll t$. In this region the hypergeometric and the interaction terms are given by
\begin{equation}\begin{split}\label{eq:Q0Qint}
    Q_0^\text{out}(z)&=-\frac{\lambda^\text{out}}{(z-1)z}+\frac{\frac{1}{4}-a_0^2}{z^2}+\frac{\frac{1}{4}-a_1^2}{(z-1)^2}\,,\\
    Q_{\text{int}}^\text{out}(z)&=+\frac{-4 z a_t^2-4 t u+4 u z+z}{4 z (t-z)^2}\,,
\end{split}\end{equation}
where 
\begin{equation}\label{eq:lambdaatainf}
    \lambda^\text{out} := -a_t^2+a_\infty ^2-a_0^2-a_1^2+u+\frac{1}{2}\,.
\end{equation}
Comparing \eqref{eq:Q0Qint} with \eqref{eq:Heun}, it is clear that the singularities at $z=0,1$ of the hypergeometric equation have the same critical exponents as those in the original Heun equation. The singularity at $z=\infty$ of the hypergeometric equation instead appears at a non-singular point of the Heun equation, as we are restricting to $z\ll t$. Moreover, the critical exponent is not that of the singularity at $z=\infty$ of the Heun equation. 

It is convenient to introduce the following notation
\begin{equation}\label{eq.operatorsnot}
    \mathcal{D}_2:=\frac{1}{w^\text{out}}\mathcal{L}_2,\quad\qquad\mathcal{D}^{\text{out},(0)}_2 :=\frac{1}{w^\text{out}}\,\mathcal{L}^{\text{out},(0)}_2\,,
\end{equation}
where we factored out 
\begin{equation}
    w^\text{out}=w^\text{out}(z):=\frac{1}{z(z-1)}\,.
\end{equation}
We also define $V_0^\text{out}=-Q_0^\text{out}/w$ and $V_\text{int}^\text{out}=-Q_{\text{int}}^\text{out}/w$. We define the bare differential operator 
\begin{equation}\label{eq:BareDiffOut}
    \widetilde{\mathcal{D}}^\text{out}_2=\frac{1}{w^\text{out}(z)}\biggl(\frac{\frac{1}{4}-a_0^2}{z^2}\,+\frac{\frac{1}{4}-a_1^2}{(z-1)^2}\biggr)\,.
\end{equation}
This operator is closely related to the leading outer hypergeometric operator
\begin{equation}
    \mathcal{D}_2^{\text{out},(0)}=\widetilde{\mathcal{D}}_2^\text{out}-\lambda^\text{out}\,,
\end{equation}
and they share eigenfunctions.

The parameters entering the Heun equation
\begin{equation}\label{eq:Pars}
    a_{t},a_\infty,a_0,a_1,u,
\end{equation}
may implicitly depend on $t$. At least initially, we will assume such dependence to be regular as $t$ approaches $\infty$. For example
\begin{equation}
    u= u_0 +\frac{u_1}{t}+\frac{u_2}{t^2}+\dots\,.
\end{equation}
In the physical problems of interest
\begin{equation}
    a_{t}=a_t(\omega)\,,\quad a_\infty=a_{\infty}(\omega),\quad a_0=a_0(\omega),\quad a_1=a_1(\omega),\quad u=u(\omega)\,,
\end{equation}
in such a way that at leading order in the large-$t$ expansion at least one of them is proportional to $\omega$, the temporal frequency of a physical fluctuation.

Under the assumption above, for any fixed $z$ within the outer region
\begin{equation}
    0\leq z\ll t\to\infty \,,
\end{equation}
the following holds
\begin{equation}
    \left|\frac{V_{\text{int}}^\text{out}(z)}{V_0^\text{out}(z)}\right|\underset{}{\ll} 1\,.
\end{equation}
Hence, in the outer region we can safely consider $V_{\text{int}}^\text{out}$ to be a small perturbation with respect to $V_0^\text{out}$ in a perturbative expansion around $t=\infty$.

To define basis elements for the functional space of solutions of the Heun problem in the outer region, we consider the following auxiliary differential equation
\begin{equation}\label{eq.tildeD2out}
    (\widetilde{\mathcal{D}}_2^\text{out}-\gamma^\text{out})\psi_\omega^\text{out}=0\,.
\end{equation}
The two local solutions of \eqref{eq.tildeD2out} at $z=0$ differ by the sign of the critical exponent $a_0$. These are given by
\begin{equation}\label{eq:SeedHyper}
    \psi_\omega^\text{out}(z)=z^{\frac{1}{2}+s_0 a_0}(1-z)^{\frac{1}{2}-a_1}\,{}_2F_1\left(\frac{1}{2}+s_0 a_0-a_1-\kappa,\frac{1}{2}+s_0 a_0-a_1+\kappa;1+2s_0 a_0;z\right),
\end{equation}
with
\begin{equation}\label{eq:KappaGamma}
    \kappa=\kappa(\omega):=\frac{1}{2} \sqrt{-1+4a_0^2+4a_1^2+4\gamma^\text{out}}\,, \qquad s_0=\pm 1\,.
\end{equation}
Without loss of generality, we define
\begin{equation}\label{eq:DicA0U00}
    \gamma^\text{out}:=-a_t^2+a_\infty ^2-a_0^2-a_1^2+u_0+\frac{1}{2}\,,
\end{equation}
then the relation between the parameters $\lambda^\text{out}$ and the parameters $\gamma^\text{out}$ is
\begin{equation}\label{eq:lambdaut}
    \lambda^\text{out}=\gamma^\text{out}+\frac{u_1}{t}+\frac{u_2}{t^2}+\dots\quad\Rightarrow\quad \lambda^\text{out}-u=\gamma^\text{out}-u_0\,,
\end{equation}
which follows from~\eqref{eq:lambdaatainf} and~\eqref{eq:DicA0U00}.

For the moment we will assume
\begin{equation}\label{eq:a0Domain}
    s_0\,\text{Re}(a_0)\geq0\,,
\end{equation}
where the sign $s_0=\pm 1$ will be selected as a function of $a_0$ in such a way that the inequality \eqref{eq:a0Domain} is obeyed. Then, in the domain
\begin{equation}
    -\frac{1}{2}<\text{Re}(a_0),\text{Re}(a_1)<\frac{1}{2}
\end{equation}
the solution in \eqref{eq:SeedHyper} is pseudo-square-integrable for both choices of $s_0$, and for any value of $\gamma^\text{out}$, since the relevant Frobenius solutions near $z=0$ and $1$ are
\begin{equation}
    \psi_\omega^\text{out}(z)\sim z^{\frac{1}{2}\pm a_0},\quad\psi_\omega^\text{out}(z)\sim \mathcal{C}_+^{(0,1)}(1-z)^{\frac{1}{2}+a_1}+\mathcal{C}_-^{(0,1)}(1-z)^{\frac{1}{2}-a_1}\,.
\end{equation}
Here, $\mathcal{C}_\pm^{(0,1)}$ are the hypergeometric connection coefficients between $z=0$ and $z=1$. The parameter $\kappa$ controls the asymptotic behaviour at $z=\infty$
\begin{equation}\label{eq.solout-zinfinity}
    \psi_\omega^\text{out}(z)\sim\mathcal{C}_+^{(0,\infty)}z^{\frac{1}{2}+\kappa}+\mathcal{C}_-^{(0,\infty)}z^{\frac{1}{2}-\kappa}\,,
\end{equation}
where $\mathcal{C}_\pm^{(0,\infty)}$ are the connection coefficients between $0$ and $\infty$.

\subsection{Outer Discrete Bases}
The local Frobenius solutions found above depend on a continuous parameter $\omega$. A discrete spectrum for $\omega$ is obtained only after imposing a second boundary condition at the other endpoint, which can be expressed as a condition on the hypergeometric connection coefficients. We first recall the standard Frobenius--Frobenius choice, which gives the untwisted $\mathrm{SL}(2,\mathbb{R})$ discrete basis. We then introduce the more general twisted, or Robin, basis. This second choice is the one naturally adapted to the large-$t$ expansion of the full Heun problem, since the information from the complementary region is encoded in the Robin parameter that will later be fixed by matching to the inner solution.

\paragraph{Discrete Spectra.}
A discrete spectrum for $\omega$ is obtained only after imposing a boundary condition for instance, at $z=1$. Such a condition may have the form
\begin{equation}
    \mathcal{C}^{(0,1)}_+-\mathcal{G}_1^\text{out}(t)\mathcal{C}_-^{(0,1)}=0\,,
\end{equation}
for some choice of function $\mathcal{G}_1^\text{out}$. We refer to the basis/spectrum defined by the conditions
\begin{equation}\label{eq:UntwistedBC}
    \mathcal{C}^{(0,1)}_+=0,\qquad\text{or}\qquad  \mathcal{C}_-^{(0,1)}=0\,,
\end{equation}
as \emph{untwisted}. Equivalently, this is defined by
\begin{equation}
    \mathcal{G}_1^\text{out}\to 0,\qquad\mathcal{G}_1^\text{out}\to\infty\,.
\end{equation}
We refer to any other discrete basis/spectrum as \emph{twisted or Robin}. Similar boundary conditions have been considered in the context of an AdS$_5$ black brane \cite{Faulkner:2009wj,Ren:2012hg}. The untwisted basis corresponds to a discrete representation of $\mathrm{SL}(2,\mathbb{R})$. Instead, Robin boundary conditions explicitly break $\mathrm{SL}(2,\mathbb{R})$.

Alternatively, the quantization conditions can be imposed at $z=\infty$. As in the previous case, a twisted discrete basis is defined by the conditions
\begin{equation}\label{eq.Goutinf}
    \mathcal{C}_+^{(0,\infty)}-\mathcal{G}_\infty^\text{out}(t)\mathcal{C}_-^{(0,\infty)} =0\,,
\end{equation}
for some choice of function $\mathcal{G}_\infty^\text{out}(t)$. The untwisted discrete basis/spectrum is obtained by the conditions
\begin{equation}\label{eq:FrobeniusOut}
    \mathcal{C}_+^{(0,\infty)}=0,\qquad\text{or}\qquad\mathcal{C}_-^{(0,\infty)}=0\,,
\end{equation}
which are the only choices not breaking $\mathrm{SL}(2,\mathbb{R})$.

\paragraph{$\mathrm{SL}(2,\mathbb{R})$ Invariant Choice of Discrete Basis.}
We first recall the standard untwisted hypergeometric basis obeying Frobenius boundary conditions at both endpoints $z=0$ and $z=\infty$. The continuous variable to discretize may be any of the continuous available parameters
\begin{equation}
    \gamma^\text{out}, \kappa, a_{\infty}, a_t,\dots\,.
\end{equation}
Depending on the specific problem one choice may be more convenient than others. In the physical problems in which we will be interested, the continuous variable to solve for will be the frequency $\omega$ of some fluctuations. As noted above, all the continuous parameters will be functions of $\omega$.

To illustrate, let us focus on $\gamma^\text{out}$ without loss of generality. Let us assume that, in the problem of interest, the dependence
\begin{equation}\label{eq:DepGammaOmega}
    \gamma^\text{out}=\gamma^\text{out}(\omega)
\end{equation}
is invertible. Under this assumption the spectrum for $\omega$ can be determined from the spectrum for $\gamma^\text{out}$. If this condition is not satisfied, for example, if $\gamma^\text{out}$ is independent of $\omega$, then another continuous variable should be chosen, e.g., $a_{\infty}$, etc, until an invertible dependence on $\omega$ is found.

The explicit expressions for the relevant connection coefficients are
\begin{equation}\begin{split}\label{eq.connection0infinity}
    \mathcal{C}^{(0,\infty)}_\pm=&\frac{\Gamma(1+2a_0)\Gamma(\pm 2\kappa)}{\Gamma\left(\pm\kappa+s_0a_0-a_1+\frac{1}{2}\right)\Gamma \left(\pm\kappa+s_0a_0+a_1+\frac{1}{2}\right)}\,.
\end{split}\end{equation}
Let us assume that we pick the Frobenius boundary condition to be $\mathcal{C}_+^{(0,\infty)}=0$. As we said, in physical setups, the parameters $\kappa,a_0,a_1$ depend on the frequency of the wave-function. We look for frequencies
\begin{equation}\label{eq:ConditionLinearL}
\omega=\omega^{(0)}_n:\qquad\kappa+s_0a_0-s_1a_1+\frac{1}{2}=-n,\qquad n\in\mathbb{Z}_{\geq 0}\,.
\end{equation}
Here, $s_1=\pm 1$ stands for poles of either the first or the second $\Gamma$-function in the denominator of~\eqref{eq.connection0infinity}.

Evaluating~\eqref{eq:SeedHyper} on these frequencies, we find
\begin{equation}\label{eq.phin}
    \phi_n^\text{out}(z)=C_n\psi_{\omega=\omega_n^{(0)}}^\text{out}(z), \qquad n\in\mathbb{Z}_{\geq 0}\,.
\end{equation}
In terms of the continuous variables $\gamma^\text{out}$ the conditions~\eqref{eq:ConditionLinearL} read
\[\gamma^\text{out}=\gamma^\text{out}(\omega_n^{(0)})=:\gamma^{\text{out},(0)}_n\,,\]
where
\begin{equation}\label{eq:UntwistedEV}
    \gamma^{\text{out},(0)}_n:=n^2+n\left(-2a_1s_1+2 a_0 s_0+1\right)-\frac{1}{2}\left(2a_0s_0+1\right)\left(2 a_1s_1-1\right)\,.
\end{equation}
The $C_n$ are complex constants that will be fixed below. 

Substituting the parameters \eqref{eq:UntwistedEV} in \eqref{eq:SeedHyper}, we find
\begin{equation}\label{eq:BasisElementsUntwisted}
    \phi_n^\text{out}(z)=C_n z^{a_0 s_0+\frac{1}{2}} (z-1)^{\frac{1}{2}- a_1} \,_2F_1\left(-n-(1-s_1)a_1,n+2 a_0 s_0-(1+s_1)a_1+1;2
   a_0 s_0+1;z\right)\,.
\end{equation}
By construction, these are eigenfunctions of the bare operator $\widetilde{\D}_2^\text{out}$ \eqref{eq:BareDiffOut}, which needs not be Hermitian. The choice of sign $s_0=\pm 1$ corresponds to picking the leading or the subleading Frobenius asymptotic solution at the endpoint $z=0$.

Assuming
\begin{equation}\label{eq.assumptiona0}
    |\text{Re}{(a_0)}|>\frac{1}{2}
\end{equation}
there is a single choice of $s_0$ for which the $\phi_n^\text{out}(z)$ are pseudo-square-integrable \eqref{eq:a0Domain}. Once such a choice of $s_0$ is picked, there are different choices of $s_1$ (depending on $a_1$) 
\begin{equation}
    s_1=\begin{cases}
        \pm 1, & -\frac{1}{2}<\text{Re}(a_{1})<\frac{1}{2},\\[0.6ex]
        -1, & \text{Re}(a_{1})\ge\frac{1}{2},\;\;2a_{1}\neq\frac{m}{2}\,, \ m=1,\dots,n-1,\\[0.6ex]
        +1, & \text{Re}(a_{1})\le-\frac{1}{2}\,,\\[0.6ex]
        +=-, & a_1=\frac{m}{2}\,,
    \end{cases}
\end{equation}
for which the following orthonormality conditions are satisfied
\begin{equation}\label{eq:Measure}
    \langle\overline{\phi_m^\text{out}},\phi_n^\text{out}\rangle:=\int_0^1 {\overline{\phi_m^\text{out}}(z)}\phi_n^\text{out}(z)\dd z=\delta_{mn}\,.
\end{equation}
Here, we have defined
\begin{equation}
    \overline{\phi_m^\text{out}}(z):=w^\text{out}(z)\phi_m^\text{out}(z)\,,
\end{equation}
and the normalization condition~\eqref{eq:Measure} fixes the complex c-number $C_n$ in \eqref{eq.phin}. Note that $\overline{\phi_m^\text{out}}$ does not denote complex conjugated. We will just refer to it as the dual wave-function to $\phi_{m}^\text{out}$. 

Using any one of these two untwisted discrete bases, we can define the inverse operator of $D^{\text{out},(0)}_2$ \eqref{eq.operatorsnot} by the equation
\begin{equation}\begin{split}
    \widehat{\mathcal{D}}^{\text{out},(0)}_{2}\cdot\left(\widehat{\mathcal{D}}_2^{\text{out},(0)}\right)^{-1}&:=\mathbb{I}\\
    &:=\sum_{n\geq 0}{|\phi_n^\text{out}\rangle\langle\overline{\phi_n^\text{out}|}}\,.
\end{split}\end{equation}
The solution is
\begin{equation}
    \left(\widehat{\mathcal{D}}_2^{\text{out},(0)}\right)^{-1}=\sum_{{n}\geq 0}\frac{|\phi_n^\text{out}\rangle\langle\overline{\phi_n^\text{out}|}}{(\gamma^{\text{out},(0)}_n-\lambda^\text{out})}\,.
\end{equation}
The identity operator $\mathbb{I}$ belongs to the subspace of pseudo-square-integrable functions, in the sense of the bilinear product defined in~\eqref{eq:Measure}, consisting of functions that satisfy Frobenius boundary conditions at $z=0$ and $z=\infty$ within a single hypergeometric patch.\footnote{The Frobenius conditions at $z=0$ and $z=\infty$ specify the relevant solution space, whereas pseudo-square-integrability itself is determined by the behaviour of the wave-functions near $z=0$ and $z=1$.}

This $\mathrm{SL}(2,\mathbb{R})$-invariant choice of discrete basis, and of corresponding gravitational functional space, is not adapted to compute subleading corrections in the large-$t$ expansion of the complete spectral problem of interest.

\paragraph{Twisted Discrete Spectrum: Explicit Breaking of $\mathrm{SL}(2,\mathbb{R})$.}
As we will show in \autoref{sec:3}, the correct choice of discrete basis $\{\phi_n^\text{out}\}$ to compute subleading large-$t$ corrections to the complete spectral problem requires imposing Robin boundary conditions at the endpoint $z=\infty$. As in the untwisted case, the twisted basis elements can be defined using \eqref{eq:SeedHyper}
\begin{equation}
    \phi_n^\text{out}=C_n\psi_{\omega=\omega_n}^\text{out}(z)\,,
\end{equation}
where
\begin{equation}
    \omega_n:\qquad \gamma^\text{out}(\omega_n)=\gamma^\text{out}_{n}\,.
\end{equation}
In particular, we note that $\gamma^\text{out}_n \neq \gamma^{\text{out},(0)}_n\,.$  Namely, they remain Gauss hypergeometric functions, but they are no longer discrete representations of $\mathrm{SL}(2,\mathbb{R})$. 

The eigenvalues $\gamma^\text{out}_n$ are solutions of Robin-like quantization condition of the form~\eqref{eq.Goutinf}
\begin{equation}\label{eq:Relation1}
    \mathcal{G}_\infty^\text{out}(t)=\frac{\mathcal{C}^{(0,\infty)}_+}{\mathcal{C}^{(0,\infty)}_-}= \frac{\Gamma (2 \kappa ) \Gamma \left(-\kappa +a_0-a_1+\frac{1}{2}\right) \Gamma\left(-\kappa +a_0+a_1+\frac{1}{2}\right)}{\Gamma (-2 \kappa ) \Gamma \left(\kappa+a_0-a_1+\frac{1}{2}\right) \Gamma \left(\kappa +a_0+a_1+\frac{1}{2}\right)}\,,
\end{equation}
where recall that $\kappa$ is a function of $\gamma^\text{out}$ \eqref{eq:KappaGamma} and the connection coefficients are those in \eqref{eq.connection0infinity}. The twisting function $\mathcal{G}_\infty^\text{out}(t)$ will be fixed by requiring the inner and outer solutions to patch smoothly. Once $\mathcal{G}_\infty^\text{out}(t)$ is fixed, one can immediately find the parameters $\gamma_n^\text{out}$ for the twisted spectrum.

The Green function of the operator $\mathcal{D}_{2}^{\text{out},(0)}$ in the functional space spanned by this basis is then
\begin{equation}\label{eq:InverseOp}
    \left(\widehat{\mathcal{D}}_2^{\text{out},(0)}\right)^{-1}:=\sum_{n\geq 0}\frac{|\phi_n^\text{out}\rangle\langle\overline{\phi_n^\text{out}|}}{(\gamma^\text{out}_n-\lambda^\text{out})}\,.
\end{equation}
where now, in contradistinction with the untwisted choice of basis
\begin{equation}\label{eq.phoutbar}
    \overline{\phi_n^\text{out}}:=w^\text{out}\sum_{m=0}^{\infty}(S^{\text{out,-1}})_n {}^m \phi^\text{out}_m\,.
\end{equation}
The overlap matrix now may be non-trivial
\begin{equation}\label{eq:OverlMatr}
    {(S^\text{out})_n}^m=\int_0^1 w^\text{out}(z)\,\phi^\text{out}_{n}(z)\,\phi^\text{out}_{m}(z)\,\dd z=\delta_{nm}+\dots\,.
\end{equation}
Here, the ellipsis denotes subleading corrections in the large-$t$ expansion. The twisted basis functions must nevertheless remain pseudo-square-integrable with respect to the same bilinear product~\eqref{eq:Measure}.

The constant $C_n$ can be fixed such that $ (S^\text{out})_n{}^{n}=1$. Computing the inverse of the operator $S^\text{out}$ can be done order by order in a large-$t$ expansion for which the twisted basis asymptotes to an $\mathrm{SL}(2,\mathbb{R})$-preserving one, or using a truncated basis for large enough values of $t$. For the purposes of this paper such computations are not necessary. Consequently, we postpone a detailed study of this overlap matrix for future work. 

The values $\gamma^\text{out}=\gamma^\text{out}_n$ are not yet the spectrum of the original Heun problem. After defining the right $\mathcal{G}_\infty^\text{out}(t)$ dictated by the problem of interest, they can be useful to define a better\footnote{With respect to the untwisted spectrum \eqref{eq:UntwistedEV}.} leading-order approximation to the Heun spectrum at large-$t$, by including small-temperature corrections that may be fractional powers of $1/t$. These corrections can be, in principle, computed using the outer/inner bilinear product, and the corresponding twisted hypergeometric basis elements, see \autoref{app:GreenF}. In \autoref{sec:3}, we will introduce an approach that trivializes the computation of such radiative corrections: in the language of this section this is nothing but a choice of twisted basis where all interaction corrections vanish.

\subsection{The Inner Hypergeometric Projection}\label{sec:AnalysisTORepeat}
The outer region expansion $t\to\infty$ is not valid uniformly at large $z$. To capture the region near the singularity at $z=\infty$, one must use a different scaling. This gives a second hypergeometric problem, dual to the outer one, i.e., the inner region problem. The complementary region of spacetime in the limit $t\to \infty$ can be found by keeping fixed the radial variable
\begin{equation}\label{eq:ChangeCoordinates}
    \zeta=\frac{z}{t}\,,
\end{equation}
and then considering the portion of spacetime defined by
\begin{equation}\label{eq:NearHorDomain}
    0\leftarrow \frac{1}{t}\ll \zeta<\infty\,.
\end{equation}
In these new coordinates the original singularities of the Heun equation map to
\begin{equation}
    z=t \longrightarrow \zeta=1\,,\qquad z=\infty \longrightarrow \zeta=\infty\,.
\end{equation}
As for the singularity at $\infty$ for the outer region, the singularities at $z=0,1$ lie outside the inner region. In the context of studying fluctuations around black hole (or black brane) backgrounds, the domain \eqref{eq:NearHorDomain} can be understood as the so-called near-horizon near-extremal region (the throat). In other problems, it may have different interpretations.

Also in this case, we can split the potential $t^2 Q$ in hypergeometric and interacting contributions
\begin{equation}
    t^2 Q=Q_0^\text{in}+Q_\text{int}^\text{in}\,,
\end{equation}
where
\begin{equation}\label{eq:Q0QintDots}\begin{split}
    Q_0^\text{in}(\zeta)&=-w^\text{in} V_0^\text{in}=-\frac{\lambda^\text{in}}{(\zeta-1)\zeta}+\frac{\frac{1}{4}-a_t^2}{(\zeta-1)^2}+\frac{a_t^2-a_\infty ^2-u}{\zeta ^2}\,, \\
    Q_\text{int}^\text{in}(\zeta)&=-w^\text{in} V_\text{int}^\text{in} =  \frac{4\zeta t\left(a_t^2-a_1^2+\lambda^\text{in}\right)+a_\infty ^2 (4-4\zeta t)+4 a_0^2 (\zeta t-1)-4 a_t^2-4\lambda^\text{in}+1}{4 \zeta ^2(\zeta t-1)^2}\,,
\end{split}\end{equation}
and 
\begin{equation}\label{eq:Rellu}\begin{split}
    w^\text{in}&=w(\zeta):=\frac{1}{\zeta (\zeta-1)}\,,\\ 
    -\lambda^\text{in}&:=u=a_t^2-a_\infty ^2+a_0^2+a_1^2+\lambda^\text{out}-\frac{1}{2}\,.
\end{split}\end{equation}
As in the outer case, we introduce the (leading) inner region hypergeometric operator
\begin{equation}\label{eq:HyperOpIn}
    \mathcal{D}_2^{\text{in},(0)}=\frac{1}{w^\text{in}}\biggl(\frac{\dd^2}{\dd\zeta^2}+Q^\text{in}_0(\zeta)\biggr)\,.
\end{equation}
For any fixed $\zeta$ within the inner region domain~\eqref{eq:NearHorDomain} it follows
\begin{equation}
    \left|\frac{V_\text{int}^\text{in}(\zeta)}{V_0^\text{in}(\zeta)}\right|\ll 1\,,
\end{equation}
which means that in the inner region we can safely consider $V_\text{int}^\text{in}$ to be a small perturbation with respect to $V_0^\text{in}$ in a perturbative expansion around $t=\infty$.

Contrary to the outer region, where a small $1/t$ expansion of $V_\text{int}^\text{out}$ is equivalent to a small $z$ expansion, the small $1/t$ expansion of $V_\text{int}^\text{in}$ is a large-$\zeta$ expansion. The reason for this is that the interaction potential in the two regions are weighted by powers of $z/t$ and $\zeta t$. On the other hand, very large values of $z$, or very small values of $\zeta$, make the large-$t$ corrections diverge in the outer and inner descriptions, respectively. This divergence of the interaction corrections is a manifestation of the non-commutativity of the two limits, $t\to\infty$ and $z\to\infty$ (resp. $t\to\infty$ and $\zeta\to0$).

We will eventually glue the outer and inner constructions at a region of small $\zeta$ and large $z$, i.e., at the matching region
\begin{equation}\label{eq.matchingsurface}
    1\ll z \ll t\to\infty, \qquad 0\leftarrow\frac{1}{t}\ll\zeta \ll 1\,,
\end{equation}
which is parametrically far from the divergences, at a distance controlled by how far $t$ is from $\infty$. 

As for the outer region \eqref{eq:BareDiffOut}, we introduce the bare differential operator to define twisted basis elements 
\begin{equation}
    \widetilde{\mathcal{D}}_2^\text{in}:=\frac{1}{w^\text{in}}\left(\frac{\dd^2}{\dd\zeta^2}+\frac{\frac{1}{4}-\kappa^2}{\zeta^2}+\frac{\frac{1}{4}-a_t^2}{(\zeta-1)^2}\right)\,.
\end{equation}
This operator is closely related to the inner region hypergeometric
\begin{equation}
    \mathcal{D}_2^{\text{in},(0)}=\widetilde{\mathcal{D}}_2^\text{in}-\lambda^\text{in}\,.
\end{equation}
In particular, they share eigenfunctions.

As for the outer region, to construct the twisted basis we start from the equation
\begin{equation}    
    \left(\widetilde{\mathcal{D}}_2^\text{in}-\gamma^\text{in}\right)\, \psi_\omega^\text{in}=0\,,
\end{equation}
where     
\begin{equation}\label{eq:SeedHyperInner}
    \psi_\omega^\text{in}(\zeta)=\left(\frac{1}{\zeta }-1\right)^{a_t+\frac{1}{2}} {\zeta}^{\frac{1}{2}\mp a_\infty}\,{}_2F_1\left(\frac{1}{2}+ \kappa+a_t\pm a_\infty,\frac{1}{2}-\kappa+a_t\pm a_\infty;\,1\pm 2a_\infty;\,\frac{1}{\zeta}\right)\,.
\end{equation}
We define
\begin{equation}\label{eq:DicA0U0}
    \gamma^\text{in}:=-u_0=-a_t^2+a_\infty ^2-a_0^2-a_1^2-\gamma^\text{out}+\frac{1}{2}\,.
\end{equation}
Here, $u_0$ denotes the leading-order term in an expansion of $u$ in $1/t$.\footnote{In particular, in \autoref{sec:3} we will show that the expansion of $u(t)$ which naturally arises in our construction -- to avoid logarithmic dependence in wave-functions -- is precisely the one determined by the Matone relations \cite{Matone:1995rx}. By construction, it is the condition imposed by fixing the monodromy around a cycle surrounding both $z=0$ and $z=1$ \cite{Lisovyy:2022flm}, as $t$ is deformed away from infinity while keeping fixed all remaining parameters.} The relation between the parameters $\lambda^\text{in}$ and the parameters $\gamma^\text{in}$ is as follows
\begin{equation}\label{eq:lambdaut2}
    \lambda^\text{in}=\gamma^\text{in}-\frac{u_1}{t}-\frac{u_2}{t^2}-\dots\quad\Rightarrow\quad \lambda^\text{in}+u=\gamma^\text{in}+u_0=0\,.
\end{equation}
This also follows from~\eqref{eq:Rellu} and~\eqref{eq:DicA0U0}.

The Frobenius solutions near $\zeta=\infty$ and $\zeta=1$ are
\begin{equation}
    \psi_\omega^\text{in}(\zeta)\sim \zeta^{\frac{1}{2}\mp a_\infty},\quad\psi_\omega^\text{in}(z)\sim \mathcal{C}_+^{(\infty,1)}(1-\zeta)^{\frac{1}{2}+a_t}+ \mathcal{C}_-^{(\infty,1)}(1-\zeta)^{\frac{1}{2}-a_t}\,.
\end{equation}
Therefore, this solution, with continuous parameter $\gamma^\text{in}$, decays at $\zeta=\infty$ and $\zeta=1$ if
\begin{equation}\label{eq:SIntebalityNearHorizonApproach}
    \pm\text{Re}(a_\infty)>0, \qquad |\text{Re}(a_t)|<\frac{1}{2}\,.
\end{equation}
If, on the other hand,
\begin{equation}
    |\text{Re}(a_t)|\geq \frac{1}{2}\,,
\end{equation}
then only one of the two local Frobenius exponents around $\zeta=1$ is decaying. The parameter $\kappa$ controls the asymptotic behaviour at $\zeta=0$
\begin{equation}\label{eq.solin-zinfinity}
    \psi_\omega^\text{in}(\zeta)\sim\mathcal{C}^{(\infty,0)}_+\zeta^{\frac{1}{2}+\kappa}+\mathcal{C}_-^{(\infty,0)}\zeta^{\frac{1}{2}-\kappa}\,.
\end{equation}
Consistently, this critical exponent matches that for the solution in the outer region at $z=\infty$ \eqref{eq.solout-zinfinity}. Note that this happens for both choices of sign in~\eqref{eq:SeedHyperInner}.

\subsection{Inner Discrete Bases}
The twisted discrete spectrum is determined by imposing the Robin boundary condition at $\zeta\simeq 0$
\begin{equation}\label{eq:Relation2}
    t^{-2\kappa}\frac{\mathcal{C}_+^{(\infty,0)}}{\mathcal{C}_-^{(\infty,0)}}=\mathcal{G}_0^\text{in}(t)\,.
\end{equation}
The continuous variable to discretize may be any of the continuous parameters at disposal
\begin{equation}
    \gamma^\text{out/in}, \kappa, a_{\infty}, a_t, \text{etc}.
\end{equation}
All of them are functions of $\omega$. By substituting the explicit expressions for the connection coefficients, we find
\begin{equation}\label{eq:RatioCoefficeintsZero}
    t^{-2\kappa}\frac{\mathcal{C}^{(\infty,0)}_+}{\mathcal{C}^{(\infty,0)}_-}=t^{-2\kappa}\frac{\Gamma(-2\kappa)\Gamma\left(\pm a_\infty+\kappa+a_t+\frac{1}{2}\right)\Gamma\left(\pm a_\infty +\kappa-a_t+\frac{1}{2}\right)}{\Gamma(2\kappa)\Gamma \left(\pm a_\infty-\kappa-a_t+\frac{1}{2}\right) \Gamma \left(\pm a_\infty-\kappa+a_t+\frac{1}{2}\right)}\,.
\end{equation}
The factor of $t^{-2 \kappa}$ comes from the change of variables $z=\zeta t$ between the two regions and the dependence on $\gamma^\text{in}$ enters through $\kappa$. 

With a judicious choice of complex normalization constant $C_n$, the corresponding basis elements
\begin{equation}
    \phi_n^\text{in}=C_n\psi_{\gamma_n}^\text{in}
\end{equation}
 are orthonormal with respect to the non-Hermitian bilinear product
\begin{equation}\label{eq:MeasureInner}
    \langle\overline{\phi_m^\text{in}},\phi_n^\text{in}\rangle:=\int_1^\infty\overline{\phi_m^\text{in}}(\zeta)\,\phi_n^\text{in}(\zeta)\dd\zeta=\delta_{mn}\,.
\end{equation}
This inner bilinear product appears to be different from \eqref{eq:Measure} but it is essentially the same, as one is recovered from the other using the inversion relation $\zeta\to1/\zeta$. As in \eqref{eq.phoutbar}, we define
\begin{equation}
    \overline{\phi_n^\text{in}}:=w^\text{in}\sum_{m=0}^{\infty}\,(S^{\text{in},-1})_n{}^m \phi^\text{in}_m\,.
\end{equation}
The overlap matrix may be non-trivial
\begin{equation}\label{eq:OverlappM}
    {(S^\text{in})_n}^m=\int_1^\infty w^\text{in}\,\phi^\text{in}_{n}(\zeta)\,\phi^\text{in}_{m}(\zeta)\,\text{d}\zeta =\delta_{nm}+\dots\,,
\end{equation}
where the ellipsis denotes subleading corrections in the large-$t$ expansion.

The constant $C_n$ can be fixed such that $ (S^\text{in})_n{}^{n}=1$. Computing the inverse of the operator $S^\text{in}$ can be done order by order in a large-$t$ expansion for which the twisted basis asymptotes to an $\mathrm{SL}(2,\mathbb{R})$-preserving one, or using a truncated basis for large enough values of $t$. Analogously to the outer case, the inner bilinear product can be used to compute radiative corrections using twisted hypergeometric basis elements in the inner space (as summarized in \autoref{app:GreenF}).\footnote{For the inner region the method is completely analogous, but using the corresponding basis, hypergeometric potential, interacting potential, and inner bilinear product~\eqref{eq:MeasureInner}.}

\subsection{Gluing Outer and Inner Problems}\label{eq:GluingSection}
Consistency in the matching region requires the outer and inner Robin
conditions to encode the same global Frobenius boundary conditions of
the Heun problem. Equivalently, the corresponding ratios of connection
coefficients must satisfy
\begin{equation}\label{eq.matchingzeroth}
    \mathcal{G}_\infty^\text{out}(t)=\mathcal{G}_0^\text{in}(t)\,.
\end{equation}
By using \eqref{eq:Relation1} and \eqref{eq:Relation2}-\eqref{eq:RatioCoefficeintsZero}, this condition can be rewritten as
\begin{equation}\begin{split}\label{eq:BasisDefinition}
    &\frac{\Gamma(2\kappa)\Gamma\left(-\kappa+a_0-a_1+\frac{1}{2}\right)\Gamma\left(-\kappa +a_0+a_1+\frac{1}{2}\right)}{\Gamma(-2\kappa)\Gamma \left(\kappa+a_0-a_1+\frac{1}{2}\right) \Gamma\left(\kappa+a_0+a_1+\frac{1}{2}\right)} \,t^{2\kappa }\\  
     &=\frac{\Gamma (-2 \kappa)\Gamma\left(\pm a_\infty+\kappa+a_t+\frac{1}{2}\right)\Gamma\left(\pm a_\infty +\kappa-a_t+\frac{1}{2}\right)}{\Gamma (2 \kappa )\Gamma\left(\pm a_\infty-\kappa-a_t+\frac{1}{2}\right) \Gamma\left(\pm a_\infty-\kappa+a_t+\frac{1}{2}\right)}\,.
\end{split}\end{equation}
To derive this expression we only considered the hypergeometric equations arising, at leading order, from the Heun equation in the inner and outer regions. Hence, we can only trust it at leading order in the small $1/t$ expansion.

We can use \eqref{eq:BasisDefinition} to find the spectrum at leading order. We restrict to modes whose frequency goes as $1/t$ in the limit $t\to\infty$,
\begin{equation}\label{eq:SmallFrequency}
    \omega \underset{t\to \infty}{=}\mathcal{O}\left(\frac{1}{t}\right).
\end{equation}
Further assuming that
\begin{equation}\label{eq:Kappa}
    \text{Re}(\kappa)<0\,,
\end{equation}
then the term on the left-hand side of \eqref{eq:BasisDefinition} vanishes for generic values of the arguments of the $\Gamma$-functions. Then, the only way to satisfy the equality is to be near poles of the $\Gamma$-functions, either in the denominator on the right side or in the numerator on the left side.\footnote{An analogous procedure applies to the alternate choice to \eqref{eq:Kappa}, $\text{Re}(\kappa)>0$. Now the poles to pick up are the ones in the numerator (resp. denominator) of the right-hand (resp. left-hand) side of \eqref{eq:BasisDefinition}.}

For black holes or black branes described by a Heun equation \eqref{eq:Heun}, $a_0$, $a_1$ and $\kappa$ do not depend on $\omega$ at leading order in $1/t$. Hence, we need to look for frequencies satisfying\footnote{In the Kerr black holes that we will study, for some eigenmodes, the magnitude $\kappa$ will happen to asymptote to a semi-integer value as well as $a_0$ and $a_1$. However, the zero coming from $\Gamma(2\kappa<0)$ in the denominator of the right-hand side of \eqref{eq:BasisDefinition} cancels a zero in the left-hand side reducing \eqref{eq:BasisDefinition} to a relation among finite quantities. The zero in the left-hand side comes from the fact that {the factor $\Gamma\left(\kappa+a_0-a_1+\frac{1}{2}\right)\Gamma\left(\kappa +a_0+a_1+\frac{1}{2}\right)$ will contribute a double zero, which due to the single pole coming from the numerator $\Gamma(2\kappa<0)$, becomes a single zero. } } 
\begin{equation}\label{eq:BasisDefinitionTot}
   \pm a_\infty-\kappa+a_t+\frac{1}{2}=-p, \qquad \text{or} \qquad \pm a_\infty-\kappa-a_t+\frac{1}{2}=-p,\qquad p\in\mathbb{Z}_{\geq 0}\,.
\end{equation}
This means that, under the ``low-frequency'' assumption \eqref{eq:SmallFrequency}, the eigenvalues are determined by the inner region through the parameters $\pm a_\infty$, $a_t$ and $\kappa$. This is consistent with the physical expectation that at small enough temperatures, eigen-fluctuations with infinitesimal frequencies~\eqref{eq:SmallFrequency} should localize near the relevant horizon where the $\mathrm{SL}(2,\mathbb{R})$ isometry is asymptotically realized at zero temperature.

To impose regularity at $\zeta=0$, we need to enforce
\begin{equation}\label{eq:SqIntegrability}
    \pm \text{Re}(a_{\infty}) >0
\end{equation}
by virtue of pseudo-square-integrability of the corresponding discrete twisted basis elements in the inner region $\psi_\omega^\text{in}$, which we recall are obtained from the master-form \eqref{eq:SeedHyperInner}. Instead, regularity at $\zeta=1$ of individual Frobenius solutions is always guaranteed, as it is always possible to solve one of the two conditions
\begin{equation}
    \text{Re}(a_t)<\frac{1}{2} \quad \text{  for }\quad  (1-\zeta)^{\frac{1}{2}- a_t}\qquad\text{or}\qquad \text{Re}(a_t)\,>\,-\frac{1}{2}\quad \text{  for }\quad  (1+\zeta)^{\frac{1}{2}- a_t}\,.
\end{equation}
These conditions guarantee orthonormality of the corresponding set of wave-functions with respect to the inner non-Hermitian bilinear product \eqref{eq:MeasureInner}. 

We conclude this section with a comment on the assumption we made on $a_0$ \eqref{eq.assumptiona0}, which singled out a choice for $s_0$. Had we chosen the opposite sign, we would have obtained an expression analogous to \eqref{eq:RatioCoefficeintsZero} but with opposite sign of $a_0$.\footnote{Both quantities are needed to compute the retarded Green function.}

\section{Disconnecting Interactions: Hypergeometric Reduction}\label{sec:3}
In the previous section, we considered, at leading order, the large-$t$ expansion of a Heun equation by splitting it into two hypergeometric problems, in the inner and outer regions. Our goal now is to compute corrections in $1/t$ to the spectrum-defining conditions~\eqref{eq:BasisDefinition}. The canonical way to tackle this problem is via Green-function computations; see \autoref{app:GreenF}. In this section, instead, we show how a unique choice of diffeomorphisms in the inner and outer regions makes it possible to project the Heun operator onto two Gauss hypergeometric operators that remain isospectral to the former. In this way, corrections to \eqref{eq:BasisDefinition} arise simply by taking these diffeomorphisms into account and requiring them to coincide in the matching region.

Using this method, we reproduce the same spectrum-defining Bethe-ansatz condition that would be obtained directly from the Heun equation by imposing Frobenius boundary conditions at $z=0$ and $z=\infty$. This shows that the eigenvalues and eigenfunctions of the Heun operator can be computed by focusing solely on the coupled system of isospectral hypergeometric problems. In particular, we will show how the spectrum of the Heun operator can be computed at the boundary of the inner region by encoding all information about the outer region in the choice of Robin boundary conditions. In black hole examples, this method provides an all-orders improvement of the standard near-horizon, near-extremal expansion. Although we will focus on the Heun case for concreteness, this method can be applied beyond this case.

\subsection{Isospectral Hypergeometric Reductions}
Consider the Heun equation \eqref{eq:Heun}
\begin{equation}\label{eq:Heun2}
    \left[\frac{\dd^2}{\dd z^2}+Q(z)\right]\Psi(z)=0,
\end{equation}
where the Schrödinger potential is given by
\begin{equation}\label{eq.schroedinger}
    Q(z)=\frac{\frac14-a_0^2}{z^2}+\frac{\frac14-a_1^2}{(z-1)^2}+\frac{\frac14-a_t^2}{(z-t)^2}-\frac{\frac{1}{2}-a_0^2-a_1^2-a_t^2+a_\infty^2+u}{z(z-1)}+\frac{u}{z(z-t)}\,.
\end{equation}
To avoid possible logarithmic corrections in $z$ when expanding closer to the matching region, we are forced to assume
\begin{equation}\label{eq:MatoneRelAssumption}
    u= u_0+\frac{u_1}{t}+\frac{u_2}{t^2}+\dots\,,
\end{equation}
where, at first and second order
\begin{equation}\label{eq:MatoneRel}
    u_0=-\frac{1}{4}-a_{\infty}^2+a_t^2+\kappa ^2, \qquad u_1 =  \frac{\left(-1-4 a_0^2+4 a_1^2+4 \kappa ^2\right) \left(-1-4 a_{\infty}^2+4 a_t^2+4\kappa ^2\right)}{32 \kappa ^2-8}\,.
\end{equation}
Higher-order terms can be computed analogously. Let us start by considering the outer region. Expanding the potential $Q(z)$ in $1/t$, we find
\begin{equation}
    Q\left(z\right)=Q_0^\text{out}(z)+\frac{1}{t}Q_1^\text{out}(z)+\frac{1}{t^2}Q_2^\text{out}(z)+\dots\,.
\end{equation}
Recall that, for the moment, we assume that $a_t$ and $a_\infty$ do not depend on $t$. 

We now introduce a generic diffeomorphism in the outer region which, at leading order, is the identity. This can be written as follows 
\begin{equation}\label{eq:zToZ}
    z=F^\text{out}[Z]:=Z+\frac1t f_1^\text{out}(Z)+\frac{1}{t^2}f_2^\text{out}(Z)+\frac{1}{t^3}f_3^\text{out}(Z)+\dots\,,
\end{equation}
where $f_j^\text{out}$ are arbitrary smooth functions. In order for the change of coordinate $z=F(Z)$ to preserve the Schrödinger form of the potential, the solution of the Heun equation in the outer region needs to be rescaled as follows
\begin{equation}\label{eq:InducedActionwave-function2}
    \Psi^\text{out}(F^\text{out}(Z))=\sqrt{(F^\text{out})'(Z)}\,\psi^\text{out}(Z)\,.
\end{equation}
Note that $\psi^\text{out}$ is, unlike $\psi_\omega^\text{out}$ introduced in \eqref{eq.tildeD2out}, a solution of the Heun equation in the new variable $Z$. Then, by inserting \eqref{eq:zToZ} in the Schrödinger potential \eqref{eq.schroedinger}, we find
\begin{equation}\label{eq:NonLinearDiff0}
    \widetilde{Q}^\text{out}\left(Z\right)=\left((F^\text{out}(Z))'\right)^2\,Q\!\left(F^\text{out}(Z)\right)+\frac{1}{2}\{F^\text{out},Z\}\,,
\end{equation}
where $\{F^\text{out},Z\}$ is the Schwarzian derivative of $F$ with respect to $Z$. Expanding \eqref{eq:NonLinearDiff0} in $1/t$, we find
\begin{equation}
    \widetilde{Q}^\text{out}\left(Z\right)=Q_0^\text{out}(Z)+\frac{1}{t}\widetilde{Q}_1^\text{out}(Z)+\frac{1}{t^2}\widetilde{Q}_2^\text{out}(Z)+\frac1{t^3}\widetilde{Q}_3^\text{out}(Z)+\dots\,.
\end{equation}
The leading term $Q_0(Z)$ is the limiting hypergeometric Schrödinger potential \eqref{eq:Q0Qint}. The higher-order terms $\widetilde{Q}_j^\text{out}$ are the perturbing potentials generated both by the large-$t$ expansion of the interacting potential $Q_\text{int}^\text{out}$ \eqref{eq:Q0Qint} and by the $t$-dependent coordinate redefinition \eqref{eq:zToZ}.

Then, we rewrite \eqref{eq:Heun2} as a perturbative expansion in $1/t$
\begin{equation}
    \left[\frac{\dd^2}{\dd Z^2}+Q_0^\text{out}(Z)+\frac{1}{t}\widetilde{Q}_1^\text{out}(Z)+\frac{1}{t^2}\widetilde{Q}_2^\text{out}(Z)+\dots\right]\Psi^\text{out}(Z)=0\,.
\end{equation}
Our goal is to show that the functions $f_j^\text{out}$ can then be chosen such that
\begin{equation}\label{eq.discinteractions}
    \widetilde{Q}_j^\text{out}=0,\qquad\forall\,j\geq 1\,.
\end{equation}
The equations for $f_j^\text{out}$ arising from this condition are of the form
\begin{equation}\label{eq.discinteractions2}
    \frac{1}{2} (f_j^\text{out})'''+2(f_j^\text{out})'Q_0^\text{out}+f_j^\text{out} (Q_0^\text{out})'=H_j^\text{out}\,,
\end{equation}
with inhomogeneities $H_j^\text{out}$ depending functionally on $Q_k^\text{out}$, with $k\leq j$, and on the $f_k$ with $k<j$. The expressions for these inhomogeneities can be determined by expanding the non-linear generating function equations or, equivalently, expanding beyond linear order \eqref{eq:NonLinearDiff0}.

Momentarily, we will derive explicit solutions of \eqref{eq.discinteractions}-\eqref{eq.discinteractions2}. By assuming that such solutions exist, and since the new coordinate $Z$ covers the outer region, we remain with a Gauss hypergeometric problem 
\begin{equation}\label{eq.hyperexact}
    \left[\frac{\dd^2}{\dd Z^2}+Q_0^\text{out}(Z)\right]\Psi^\text{out}(Z)=0\,.
\end{equation}
All the interactions have been disconnected \eqref{eq.discinteractions} by the choice of diffeomorphism \eqref{eq:zToZ}. As the implemented diffeomorphism does not change the spectrum, i.e., it is \emph{isospectral}, the discrete spectrum of the hypergeometric operator \eqref{eq.hyperexact}, with appropriate twisted boundary conditions at $Z=\infty$ and Frobenius boundary condition at $Z=0$,  matches, by definition, the spectrum of the undeformed Heun operator with Frobenius boundary conditions at both endpoints. 

An analogous condition can be introduced to define the isospectral diffeomorphism that disconnects interactions in the inner region
\begin{equation}\label{eq:zToWP}
    z=F^\text{in}[\wp]=t\zeta,\qquad\zeta=F^\text{in}[\wp]:=\wp+\frac{1}{t}f_1^\text{in}(\wp)+\frac{1}{t^2}f_2^\text{in}(\wp)+\frac{1}{t^3}f_3^\text{in}(\wp)+\dots\,,
\end{equation}
giving rise to the following potential
\begin{equation}\label{eq:NonLinearDiffInner}
    \widetilde{Q}^\text{in}\left(\wp\right)=\left((F^\text{in}(\wp))'\right)^2Q\left(F^\text{in}(\wp)\right)+\frac{1}{2}\{F^\text{in},\wp\}\,.
\end{equation}
As in \eqref{eq.discinteractions}-\eqref{eq.discinteractions2}, the functions $f_j^\text{in}$ can be chosen to disconnect all interactions. Hence, the resulting hypergeometric problem with Robin boundary conditions at $\wp=0$ and Frobenius at $\wp=\infty$ is equivalent to the original Heun problem with Frobenius boundary conditions at both endpoints.

The effects of the interactions $Q_\text{int}^\text{out}$ and $Q_\text{int}^\text{in}$ in both patches remain stored in the choice of Robin boundary conditions, i.e., in the choice of, respectively, $\mathcal{G}_\infty^\text{out}(t)$ \eqref{eq:Relation1} and $\mathcal{G}_0^\text{in}(t)$ \eqref{eq:Relation2}. The diffeomorphism introduces $1/t$-corrections in the functions determining the Robin boundary conditions, which we analyse below. We will show that the spectrum-generating condition we obtain reproduces the next-to-leading results originally obtained in \cite{Bonelli:2021uvf}, and later verified in \cite{Lisovyy:2022flm}.

\subsection{Explicit Construction of the Diffeomorphisms}
We now construct the outer and inner diffeomorphisms explicitly, presenting the first non-trivial terms in their large-$t$ expansions.

\paragraph{The Outer Diffeomorphism.}
We now provide an explicit solution for the first few terms in \eqref{eq.discinteractions}-\eqref{eq.discinteractions2}. Starting from $j=1$, the condition $\widetilde{Q}_1^\text{out}=0$ gives the Fuchsian differential equation of third order
\begin{equation}\label{eq:DiffEqFlucConc}
    \left(\frac{1}{2}\partial_Z^3+2Q_0^\text{out}(Z)\partial_Z+(Q_0^\text{out}(Z))'\right)f_1^\text{out}(Z)=\frac{u_0}{Z}+\frac{u_1}{Z(Z-1)}\,,
\end{equation}
The most general solution to the inhomogeneous equation is
\begin{equation}
    f_1^\text{out}(Z)=f_{1,\text{p}}^\text{out}(Z)+f_{1,\text{q}}^\text{out}(Z),
\end{equation}
where $f_{1,\text{p}}^\text{out}$ is any particular solution of the inhomogeneous equation and $f_{1,\text{q}}^\text{out}$ is a solution of the homogeneous equation. Analyticity forces the choice $f_{1,\text{q}}^\text{out}=0$.

At large $Z\to\infty$, the leading-order hypergeometric potential \eqref{eq:Q0Qint} becomes
\begin{equation}
    Q_0(Z)=\frac{\frac{1}{4}-\kappa^2}{Z^2}+\mathcal{O}(Z^{-3})\,.
\end{equation}
The choices of $u_0,u_1,\dots$ in \eqref{eq:MatoneRel} are precisely those that ensure that logarithmic contributions in $Z$ are avoided at every subleading order in the large-$Z$ expansion. It then follows that there is a unique solution, with the following asymptotic expansion up to first subleading order at large $Z$
\begin{equation}\label{eq.soldiffeoout1}
    f_{1,\text{p}}^\text{out}(Z)\underset{Z\to \infty}{\sim}A_{1,2} Z^2+A_{1,1}Z\,,
\end{equation}
with
\begin{equation}\label{eq.soldiffeoout2}
    A_{1,2}=\frac{-1-4a_\infty^2+4a_t^2+4\kappa^2}{(2-8\kappa^2)},\qquad A_{1,1}=\frac{1+4a_\infty^2-4a_t^2-4\kappa^2}{(2-8\kappa^2)}\,.
\end{equation}
Remarkably, this solution is exact in $Z$. By this we mean that in an expansion around either $Z=0$ or $Z=\infty$, the relevant perturbative approximation truncates to an exact solution. Hence, we can write
\begin{equation}\label{eq.soldiffeoout}
f_1^\text{out}=f_\text{p}^\text{out}(Z)=A_{1,2}Z^2+A_{1,1}Z\,,
\end{equation}
where we have removed the asymptotic symbol.

Analogous truncations happen at each higher order in the $1/t$-expansion. For instance, the series truncates at order $Z^3$ starting at order $Z$ (expanding around $Z=0$), or equivalently, it truncates at order $Z$ starting at order $Z^3$ (expanding around $Z=\infty$):
\begin{equation}\label{eq.soldiffeooutsec}
    f_2^\text{out}=A_{2,3}Z^3+A_{2,2}Z^2+A_{2,1}Z.
\end{equation}
We conjecture that the pattern continues at higher orders. Higher-order corrections will be studied elsewhere.

\paragraph{The Inner Diffeomorphism.}
In the inner coordinate $\zeta=z/t$, the diffeomorphism
\eqref{eq:zToWP} is determined directly from the non-linear equation
\eqref{eq:NonLinearDiffInner}. Expanding at fixed $\wp$ gives, at first order,
\begin{equation}\label{eq.soldiffeoin}
    f_1^\text{in}(\wp)=B_{1,1}\wp+B_{1,0}\,,
\end{equation}
with
\begin{equation}
    B_{1,1}=\frac{-1-4a_0^2+4a_1^2+4\kappa^2}{(2-8\kappa^2)},\qquad B_{1,0}=\frac{1+4a_0^2-4a_1^2-4\kappa^2}{(2-8\kappa^2)}\,.
\end{equation}
This solution is exact in the variable $\wp$. Similar truncations happen at higher orders in the large-$t$ expansion, e.g., at second order we find
\begin{equation}\label{eq.soldiffeoinsec}
    f_2^\text{in}(\wp)=  B_{2,1}\wp +B_{2,0}+\frac{B_{2,-1}}{\wp}\,.
\end{equation}
As for the outer region, we conjecture that the pattern continues.

\subsection{Corrections to the Spectrum-Generating Conditions}
The higher-order spectrum-generating condition follows by matching the
outer and inner transformations in the overlap region. Smooth matching
requires
\begin{equation}\label{eq:fMatch}
    F^\text{out}=z=t F^\text{in}\,,
\end{equation}
which fixes the relation between the hypergeometric coordinates
\begin{equation}\label{eq:ChangeZzeta}
    Z=t\left(1+\frac{\Delta C_0}{t}+\mathcal{O}\left(\frac{1}{t^2}\right)\right)\,(\wp+\wp_0)+(\dots)\,.
\end{equation}
Here, $\wp_0$ is a constant independent of $t$, $\wp$, and $Z$, and the ellipsis denotes subleading corrections in the matching asymptotic expansion. Note that $\wp_0$ is a translation and $\Delta C_0$ is a dilatation. Since $\wp_0$ can be absorbed by a constant shift $\wp\to\wp_\text{new}-\wp_0$, the translation has no effect on the ratios between connection coefficients. By contrast, the dilatation $\Delta C_0$ has a very important physical meaning, as we will show in \autoref{sec.3.5}. It is fixed by the smoothness condition \eqref{eq:fMatch} and it depends on the parameters of the Heun equation $\Delta C_0=\Delta C_0(a_0,a_1,u,\dots)$.

To compute the transcendental quantization condition we just repeat the same analysis presented in \autoref{sec:AnalysisTORepeat}. However, this time we have to account for the choices of diffeomorphisms
\begin{equation}
    z\to Z\,, \qquad \zeta \to \wp_{\text{new}}-\wp_0\,,\qquad t\to t\,\left[1+\frac{\Delta C_0}{t}\,+\dots\right]\,.
\end{equation}
By incorporating these corrections into the leading-order analysis, the matching condition still reads as in \eqref{eq.matchingzeroth}, that is
\begin{equation}
    \mathcal{G}_0^\text{in}=\mathcal{G}_\infty^\text{out}\,.
\end{equation}
However, the inner Robin boundary conditions include the change of coordinates between $\wp$ and $Z$
\begin{equation}\label{eq:Relation3}
    \mathcal{G}_0^\text{in}(t)=t^{-2\kappa }\left[1-\frac{2\kappa\, \Delta C_0}{t}+\mathcal{O}(t^{-2})
\right]\,\frac{\mathcal{C}_0^{(\infty,0)}}{\mathcal{C}_-^{(\infty,0)}} \, 
\end{equation}
instead of \eqref{eq:Relation2}. Thus, the condition $\Delta C_0\neq 0$ deforms the Robin boundary conditions at order $1/t$. In the following section we will compute $\Delta C_0$ (and $\wp_0$) using \eqref{eq:ChangeZzeta}, and compare the result with the instanton-counting prediction \cite{Bonelli:2021uvf,Arnaudo:2024sen}.

In summary, the spectrum is now defined by the transcendental equations
\begin{equation}\begin{split}\label{eq:BasisDefinitionCompletion}
    &\frac{\Gamma(2\kappa )\Gamma\left(-\kappa+a_0-a_1+\frac{1}{2}\right) \Gamma\left(-\kappa +a_0+a_1+\frac{1}{2}\right)}{\Gamma (-2\kappa)\Gamma\left(\kappa+a_0-a_1+\frac{1}{2}\right) \Gamma \left(\kappa +a_0+a_1+\frac{1}{2}\right)}\\  
   =&t^{-2\kappa }\left[1-\frac{2\kappa\Delta C_0}{t}+\mathcal{O}(t^{-2})\right]\frac{\Gamma (-2 \kappa ) \Gamma \left(\pm a_\infty +\kappa+a_t+\frac{1}{2}\right) \Gamma\left(\pm a_\infty +\kappa-a_t+\frac{1}{2}\right)}{\Gamma (2 \kappa ) \Gamma \left(
    \pm a_\infty-\kappa-a_t+\frac{1}{2}\right) \Gamma \left(\pm a_\infty -\kappa+a_t+\frac{1}{2}\right)}\,,
\end{split}\end{equation}
which modifies \eqref{eq:BasisDefinition} beyond leading order. As the interactions have been isospectrally disconnected, there is no need to compute further radiative corrections. 

As discussed at the end of \autoref{sec:2}, one can consider a version of this expression with the opposite sign of $a_0$. In specific physical problems, by combining the two expressions, one can obtain the retarded Green function for the full geometry at any order in $1/t$, and computed at the boundary of the inner region.

\subsection{Deriving $\Delta C_0$: Comparison with Instanton Computation }\label{sec.3.5}
To compute $\Delta C_0$ in \eqref{eq:Relation3} the procedure is as follows. First, we solve for $F^\text{out}$ and $F^\text{in}$, in the $t\to\infty$ limit, keeping $Z$ and $\wp$ fixed, respectively. To achieve this, let us recall that the potentials in the new variables are given by \eqref{eq:NonLinearDiff0} and \eqref{eq:NonLinearDiffInner}. By requiring that the interactions are disconnected, see for instance \eqref{eq.discinteractions} in the outer region, we find the following equations defining $F^\text{out}$ and $F^\text{in}$
\begin{equation}\begin{split}\label{eq:SecondEqQ0}
    Q_0^\text{out}\left(Z\right)&=((F^\text{out}(Z))')^2Q(z)+\frac{1}{2}\{F^\text{out},Z\}\,,\\ 
    Q_0^\text{in}(\wp)&=((F^\text{in}(\wp))')^2Q\!\left(z\right)+\frac{1}{2}\{F^\text{in},\wp\}\,.
\end{split}\end{equation}
We need to solve these equations order-by-order in the $1/t$-expansion, followed by an expansion in $Z$ and $\wp$.

The explicit solution of \eqref{eq:SecondEqQ0} is determined by the expansion in $1/t$ for the accessory parameter $u(t)$.\footnote{The first term of this expansion has been determined in \eqref{eq:MatoneRel}. Higher-order terms have been fixed to avoid logarithmic dependence in $z$ when expanding around the matching region \eqref{eq:MatoneRel}. Those potential logarithmic terms come from inflows in $t$ that do not preserve the critical exponent at the matching region.} The first two terms in the expansion have been derived in \eqref{eq:MatoneRel}. Overall, we find
\begin{equation}
    u=-\frac{1}{4}-a_\infty^2+a_t^2+\kappa^2+\frac{\left(-1-4a_0^2+4a_1^2+4\kappa^2\right)\left(-1-4a_\infty^2+4a_t^2+4\kappa^2\right)}{t\left(-8+32\kappa^2\right)}+\mathcal{O}\left(\frac{1}{t^2}\right)\,.
\end{equation}
The first terms can be recognized as reproducing the Matone relation \cite{Matone:1995rx}, with the critical exponent at the matching region $\kappa$ playing the role of the Coulomb branch parameter $a$.

We have already found, by inserting the expansion of $u(t)$, solutions at first and second order for the diffeomorphisms \eqref{eq:SecondEqQ0} in \eqref{eq.soldiffeoout}-\eqref{eq.soldiffeooutsec} for the outer region, and \eqref{eq.soldiffeoin}-\eqref{eq.soldiffeoinsec} for the inner region. This allows us to write the solution in the outer region
\begin{equation}\begin{split}\label{eq:Foutexp}
    F^\text{out}&=Z+ \frac{A_{1,2}Z^2+A_{1,1}Z}{t}+\dots+\frac{A_{n-1,n} Z^n+\dots\,A_{n-1,1}Z^{1}}{t^{n-1}}+\dots\\
    &=\left(1+\frac{1+4a_\infty^2-4a_t^2-4\kappa^2}{(2-8\kappa^2)t}\right)Z+\frac{(-1-4a_\infty^2+4a_t^2+4\kappa^2)Z^2}{(2-8\kappa^2)t}+\mathcal{O}\left(\frac{1}{t^2}\right)\,,
\end{split}\end{equation}
and in the inner region
\begin{equation}\begin{split}\label{eq:Finexp}
    F^\text{in}&=\wp+\frac{B_{1,1}\wp+B_{1,0}}{t}+\dots\,+\frac{B_{n+1,1}\wp\dots+\frac{B_{n+1,-n}}{\wp^n}}{t^{n+1}}+\dots\\
    &=\frac{1+4a_0^2-4a_1^2-4\kappa^2}{(2-8\kappa^2)t}+\left(1+\frac{-1-4a_0^2+4a_1^2+4\kappa^2}{(2-8\kappa^2)t}\right)\wp +\mathcal{O}\left(\frac{1}{t^2}\right)\,.
\end{split}\end{equation}
Higher-order coefficients can be determined by substituting the ansätze into the equations \eqref{eq:SecondEqQ0}, and expanding them at large $t$. Subsequently, they need to be expanded in either small $Z$ or large $\wp$. 

Second, we need to find the unique transition map between the hypergeometric coordinates $Z=Z[\wp]$ such that the outer and inner diffeomorphisms become functionally equal up to suppressed terms in the matching region
\begin{equation}\label{eq:MatchingIsospectral}
    F^\text{out}[Z[\wp]]-tF^\text{in}[\wp]=\mathcal{O}(\wp^2)+\mathcal{O}\left(\frac{1}{t^2}\right)\,.
\end{equation}
This expression can, in principle, be generalized to the desired order in the $1/t$ expansion. By imposing \eqref{eq:MatchingIsospectral} we find\footnote{Note that terms like $\frac{\text{const}}{\zeta t}$ are suppressed in this expansion, despite the fact $\zeta$ is in the denominator.}
\begin{equation}\begin{split}
    \frac{Z}{t}=&\left(1+\frac{\Delta C_{0}}{t}\right)\left(\wp+\wp_0\right)+\frac{b_0+ b_1 \wp}{t^2}+\mathcal{O}(\wp^2)+\mathcal{O}\left(\frac{1}{t^3}\right)\,,
\end{split}\end{equation}
where
\begin{equation}\begin{split}
    \wp_0&=\frac{1-4\kappa^2+4a_0^2-4a_1^2}{2-8\kappa^2},\qquad b_0= \frac{\left(1+4a_0^2-4a_1^2-4\kappa ^2\right)^2\left(-1+4 a_\infty ^2-4 a_t^2+4\kappa ^2\right)}{8\left(1-4 \kappa ^2\right)^3}+\dots\,,\\ b_1&=\frac{1}{8\left(1-4\kappa^2\right)^4}\left(1-4\kappa^2-4a_t^2+4a_\infty^2\right)\Big[8\left(1-4\kappa^2\right)^2 a_1^2+\left(1-4\kappa^2\right)^2\left(-1+4\kappa^2+4a_t^2-4a_\infty^2\right) \\
    &+16a_0^4\left(1-4\kappa^2-12a_t^2+12a_\infty^2\right)+16a_1^4\left(1-4\kappa^2-12a_t^2+12a_\infty^2\right) \\ &-8a_0^2\left(\left(1-4\kappa^2\right)^2+4a_1^2\left(1-4\kappa^2-12a_t^2+12a_\infty^2\right)\right)\Big]+\dots\,,
\end{split}
\end{equation}
The ellipsis denotes contributions at order $\mathcal{O}(1/t^2)$ not appearing in \eqref{eq:Foutexp}-\eqref{eq:Finexp}. More importantly, \eqref{eq:MatchingIsospectral} also determines
\begin{equation}\label{eq:ComputedDeltaC0}
    \Delta C_{0}=-\frac{1}{2}-\frac{8(a_0-a_1)(a_0+a_1)(a_t^2-a_\infty^2)}{(1-4\kappa^2)^2}\,.
\end{equation}
This expression matches the answer coming from the instanton analysis of \cite{Bonelli:2021uvf}. Indeed, $\Delta C_0$ satisfies
\begin{equation}
    \Delta C_0=\frac{1}{\kappa}\,\partial_\kappa\left(\frac{\left(\frac{1}{4}-\kappa^2+a_0^2-a_1^2\right)\left(\frac{1}{4}-\kappa^2-a_t^2+a_\infty^2\right)}{1-4\kappa^2}\right)\,.
\end{equation}
The term in brackets is precisely the leading term of the Nekrasov--Shatashvili free energy expanded in $1/t$.\footnote{This can be explicitly checked, for instance, by comparing with equation (30) of \cite{Arnaudo:2024sen}.} We leave the analytic resolution of~\eqref{eq:MatchingIsospectral} at higher orders in the large-$t$ expansion for future work. 

Our approach offers a perspective different from those in \cite{Bonelli:2021uvf,Lisovyy:2022flm} regarding the computation of connection coefficients. In a sense, it geometrizes the problem of computing perturbative corrections: computing them is equivalent to determining the diffeomorphisms $F^\text{out}$ and $F^\text{in}$ and then solving for the relation $Z=Z[\wp]$ that guarantees their smooth transition across the matching region. This procedure avoids the need to compute integrals and Born series expansions outlined in \autoref{app:GreenF}.

\subsection{Projection of the Wave-Function}\label{eq:Projection}
The complete wave-function can be reconstructed, in the inner and outer regions, using the inverse maps of \eqref{eq:Foutexp}-\eqref{eq:Finexp}, and the transformation law of the Schrödinger wave-function \eqref{eq:InducedActionwave-function2}. For instance, in the outer region at linear order in the $1/t$ expansion 
\begin{equation}
    \Psi^\text{out}(z)=\psi^\text{out}(z)-\frac{1}{t}\left(f_1^\text{out}\partial_z-\frac{1}{2}(f_1^\text{out})'\right)\psi^\text{out}(z)+\mathcal{O}\left(\frac{1}{t^2}\right)\,.
\end{equation}
Similarly, we can obtain $\Psi^\text{in}$ in the inner region. Beyond linear order, the same procedure applies once $F^\text{out}$ and $F^\text{in}$ have been computed to desired order in the $1/t$ expansion.

In particular, $\psi^\text{out}$ and $\psi^\text{in}$ are hypergeometric functions and the inverse diffeomorphism action is given by polynomials in $Z$ or $\wp$ times derivatives in $Z$ or $\wp$. Hence, by virtue of the truncation property \eqref{eq:Foutexp}-\eqref{eq:Finexp}, the explicit expressions for $\Psi^\text{out}(z)$ and $\Psi^\text{in}(z)$ are infinite linear combinations of either outer or inner Gauss hypergeometric functions, with coefficients depending on $t$. Namely, $\Psi^\text{out}(z)$ and $\Psi^\text{in}(z)$ can be written in terms of blocks of the form ${}_2 F_1(a,b,c;z)$ obtained from the hypergeometric functions $\psi^\text{out}$ and $\psi^\text{in}$ via integer shifts in their parameters.

This conclusion follows from the contiguous relation, which describes the behaviour of hypergeometric functions under multiplication by polynomials
\begin{equation}
    z\,{}_2F_1(a,b;c;z)=\frac{c-1}{a-b}\left[{}_2F_1(a-1,b;c-1;z)-{}_2F_1(a,b-1;c-1;z)
\right]\,,
\end{equation}
or under the action of derivatives
\begin{equation}
    \partial_z^m\,{}_2F_1(a,b;c;z)=\frac{(a)_m(b)_m}{(c)_m}{}_2F_1(a+m,b+m;c+m;z)\,.
\end{equation}
Therefore, for any polynomial $P(z)$ the following holds
\begin{equation}
    P(z)\,\partial_z^m\,{}_2F_1(a,b;c;z)=\sum_j A_j\,{}_2F_1(a+r_j,b+s_j;c+\ell_j;z)
\end{equation}
where $r_j,s_j,\ell_j\in\mathbb{Z}$.  The coefficients \(A_j\) are rational
functions of the parameters $a,b,c$ and of the coefficients of $P(z)$.

The conclusions of the Born series approach in \autoref{app:GreenF} strongly suggest that the integer-shifted hypergeometric functions correspond to the basis of pseudo-square-integrable states that we have defined before, that is either
\begin{equation}
    |\phi_n^\text{out}\rangle,\qquad \text{or}\qquad|\phi_n^\text{in}\rangle\,,
\end{equation}
depending on whether one is working in the outer or inner patch. The overlap matrices $\bigl(S^\text{out/in}\bigr)_{n}{}^{m}$ can be straightforwardly recovered from the wave-function by computing the integrals in~\eqref{eq:OverlMatr} and~\eqref{eq:OverlappM}. The details of such relations will be addressed elsewhere.

\section{Schwarzian Modes for Kerr Black Holes}\label{sec.4}
We now study the gravitational functional space of fluctuations $\mathcal{H}$. As explained before, the basis elements are defined patchwise
\begin{equation}
    |\Phi_n\rangle=\left\{\begin{array}{cc}
        |\phi_n^\text{out}\rangle & \qquad\text{outer region,} \\
        |\phi_n^\text{in}\rangle & \qquad\text{inner region.}
    \end{array}\right.
\end{equation}
The outer and inner parts were introduced in \eqref{eq:Measure} and \eqref{eq:MeasureInner}. We then define the gravitational functional space of linear fluctuations as follows
\begin{equation}
    \mathcal{H}(t) =\bigg\{|\Phi\rangle:\quad|\Phi\rangle=\sum_n  c_n(t)|\Phi_n\rangle\bigg\}\,.
\end{equation}
Our goal is to identify the maximal subspace $\mathcal{H}_\text{wzm}(t)\subset\mathcal{H}(t)$ characterized by states whose frequency vanishes at $t=\infty$. Namely, these are elements in $\mathcal{H}$ which become zero modes at $t=\infty$. 

In this section, we will only exploit the approach presented in \autoref{sec:2} to obtain the leading asymptotic form of $\mathcal{H}_\text{wzm}$ at $t\to\infty$. Exploiting the method described in \autoref{sec:3} we could, in principle, reconstruct all small-temperature corrections to $|\Phi_n\rangle$ starting from $t=\infty$. Moreover, although we work in Lorentzian signature, the analysis of the functional space is really agnostic about the signature of time.\footnote{In the Lorentzian case, we assume that the time integration domain entering the definition of the path integral is finite.} A posteriori, we will see that the modes we will find in $\mathcal{H}_\text{wzm}(t)$ naturally select one such signature. In particular, we will show that, after Wick rotation to Euclidean signature and assuming KMS conditions, they become stable Matsubara modes.\footnote{That said, we do not exclude the possibility of other Lorentzian would-be zero modes that do not transform into Matsubara modes in Euclidean signature. This problem will be addressed in future work.}

At leading order, $\mathcal{H}_\text{wzm}$ is determined by the spectrum via equation \eqref{eq:BasisDefinition}. We will check that $\mathcal{H}_\text{wzm}$ includes the expected tensor (spin 2) and vector zero modes (spin 1), which have been widely studied in the AdS$_2$ near-horizon region \cite{Iliesiu:2022onk}. This analysis explicitly shows how the Schwarzian functional space emerges both in Lorentzian and Euclidean signature from the complete functional space of fluctuations $\mathcal{H}$, at leading order in a small-temperature expansion.\footnote{Note that the analysis of \cite{Arnaudo:2024bbd,Arnaudo:2024rhv} used the Gelfand-Yaglom approach to compute one-loop determinants, or equivalently as a generalization of the DHS approach \cite{Denef:2009kn}. This approach did not provide detailed information about the functional space of states (wave-functions) over which those determinants are computed. The authors of \cite{Arnaudo:2024bbd,Arnaudo:2024rhv} reproduced the correction associated to the Schwarzian at first order in the small-temperature expansion, using as an input their expected relation to Matsubara modes. This relation is derived in our approach.} As background geometries we will use near-extremal four-dimensional Kerr black holes with either positive or negative cosmological constant.\footnote{For Kerr AdS$_4$ black holes, the AdS boundary is not a singularity of the Heun equation. Hence, one of the two singularities of the Heun equation with Frobenius boundary conditions needs to be such that the AdS boundary is within the radius of convergence of the local solutions around such singularity. However, it has been argued in \cite{Arnaudo:2024bbd} that, in the near-extremal regime, the quantization condition can be studied by considering the connection coefficients between the horizon and the other Frobenius singularity.} In particular, our approach confirms the presence of contributions from vector zero modes both in AdS and dS cases. The results in this section complement the numerical results of \cite{Kolanowski:2024} for near-extremal Kerr AdS$_4$ black holes (see also related work \cite{Rakic:2023vhv,Mariani:2025hee,PandoZayas:2026vbg}).

Our approach not only allows us to reconstruct the variations of these modes and their spectrum at any order in a small-temperature expansion, but also reveals the presence of additional competing low-temperature modes. These terms may spoil the dominance of the Schwarzian contribution at leading order in the small-temperature expansion. Such competition will be studied elsewhere.

To illustrate that the leading small-temperature asymptotic behaviour of the emergent light modes is universal and robust under the ambiguity in the choice of the defining basis $\lvert\Phi_n\rangle$, we will consider three alternative parametrizations of our radial coordinate $z$ and parameter $t$ in terms of the variables $z_\text{there}$ and $t_\text{there}$ used in \cite{Arnaudo:2025btb}. We will show that the defining properties of the emergent light modes are essentially insensitive to these different choices of basis, provided that, in the low-temperature limit, the event horizon asymptotes to one of the two endpoints $z=\{0,\infty\}$ at which the Frobenius boundary conditions defining the basis are imposed. Without loss of generality, we fix that endpoint to be $z=\infty$.

More specifically, our endpoint $z=0$ may be associated with either the unphysical horizon or the cosmological horizon, whereas $z=\infty$ may be associated with either the inner horizon or the event horizon. These different identifications correspond to distinct choices of basis ${\lvert\Phi_n\rangle}$. Nevertheless, as long as the event horizon approaches $z=\infty$ order by order in the small-temperature expansion, both the leading asymptotic behaviour and the characteristic properties of the emergent light modes that we will study remain universal. The low-temperature light sector is therefore robust under the corresponding changes of basis in the space of gravitational fluctuations over which the fluctuation path integral is evaluated.

\paragraph{Conventions.}
We begin by translating the conventions of \cite{Arnaudo:2024bbd,Arnaudo:2025btb} for the near-cold limit of a dS$_4$ Kerr black hole and a near-extremal Kerr AdS$_4$ into our notation. As explained before, to illustrate robustness under different choices of basis, we will work simultaneously with three independent sets of identifications between our radial coordinates $z$ and parameter $t$ and those used in~\cite{Arnaudo:2025btb}. These are 
\begin{equation}\begin{split}\label{eq:ChoicesRadialCoordinates}
    z&=\begin{cases}\qquad \frac{1}{z_{\text{there}}} \\\quad\,\,\frac{z_\text{there}-1}{z_\text{there}} \\ \quad \frac{z_\text{there}-1}{z_\text{there}-t_\text{there}}\end{cases}\,, \qquad\qquad  t=\begin{cases}\qquad \frac{1}{t_{\text{there}}}\\
    \quad\,\,\frac{t_\text{there}-1}{t_\text{there}} \\\qquad  \frac{1}{t_\text{there}}\end{cases}\,,
\end{split}\end{equation}
where
\begin{equation}
    z_\text{there}=\frac{(r-R_i)(R_+-R_-)}{(r-R_-)(R_+-R_i)},\qquad {t_{\text{there}}}=\frac{(R_h-R_i)(R_+-R_-)}{(R_h-R_-)(R_+-R_i)}\,.
\end{equation}
For the dS case, $R_h$ is the event horizon, $R_i$ is the inner horizon, $R_+$ is the cosmological horizon while $R_-<0$. The relation between the singularities of the potential $Q(z)$ and the horizons is as follows
\begin{equation}
    r=R_\mp\mapsto z=0,\qquad r=R_\pm \mapsto z=1,\qquad r=R_h\mapsto z=t,\qquad r=R_i\mapsto {z}=\infty\,,
\end{equation}
where the upper/lower choice of sign denotes the first/second choice in~\eqref{eq:ChoicesRadialCoordinates}. The third choice corresponds to
\begin{equation}
    r=R_+\mapsto z=0 ,\qquad r=R_- \mapsto z=1,\qquad r=R_i\mapsto z=t,\qquad r=R_h\mapsto {z}=\infty\,.
\end{equation}
This choice imposes Frobenius pseudo-square-integrability boundary conditions at the event and cosmological horizons and it is the standard choice in the literature \cite{Berti:2009kk}. The $R_\pm$ are given by
\begin{equation}\begin{split}
    R_-=&-R_+-R_h-R_i\,,\\
    R_+=&-\frac{R_h+R_i}{2}+\frac{1}{2}\sqrt{\frac{12}{\Lambda}-\left(4a_\text{BH}^{2}+3R_h^{2}+2R_hR_i+3R_i^{2}\right)}\,,
\end{split}\end{equation}
where
\begin{equation}
    a_\text{BH}^{2}=\frac{R_hR_i\left(-\Lambda R_h^{2}-\Lambda R_hR_i-\Lambda R_i^{2}+3\right)}{\Lambda R_hR_i+3}
\end{equation}
is the Kerr rotation parameter.

The temporal ($t_L$) and angular dependence of the Teukolsky modes \cite{Teukolsky:1972my,Arnaudo:2024rhv} is\footnote{We will not be concerned with the angular dependence $\Theta(\theta)$ which has been studied for instance in \cite{Arnaudo:2025btb,suzuki:1998vy,Novaes:2018fry}.} 
\begin{equation}
    \Phi\sim e^{-\ii \omega t_{L}+i m\phi}\, {\Psi}(z)\,\Theta(\theta)\,.
\end{equation}
The differential equation for the radial dependence $\Psi(z)$ is a Heun differential equation \cite{Teukolsky:1972my,Batic:2007it}. The critical exponents $a_A$ of this differential equation, where $A$ ranges over $\{\mp,\pm,h,i\}$ or $\{+,-,i,h\}$ and labels the four regular singularities, depend on the following combination of the frequency of the fluctuation $\omega$ and the angular velocities $\Omega_A$ of the black hole horizons
\begin{equation}
    \omega_A=\omega-m\Omega_A,\qquad\Omega_A=\frac{a_\text{BH}}{a_\text{BH}^2+R_A^2}\,.
\end{equation}
Explicitly
\begin{equation}
    a_A=\frac{\mathfrak{s}}{2}+\frac{\ii }{2 \kappa_A}\left(\omega-m\Omega_A\right)\,,
\end{equation}
or, equivalently,
\begin{equation}\begin{split}\label{Eq:parameters}
    a_0&=\frac{\mathfrak{s}}{2}+\frac{\ii (\omega-m\Omega_\mp)}{2 \kappa_\mp},\qquad a_1=\frac{\mathfrak{s}}{2}+\frac{\ii (\omega-m\Omega_\pm)}{2 \kappa_\pm}\,,
    \\a_t&=\frac{\mathfrak{s}}{2}+\frac{\ii (\omega-m\Omega_h)}{2 \kappa_h},\qquad a_\infty=\frac{\mathfrak{s}}{2}+\frac{\ii (\omega-m\Omega_i)}{2 \kappa_i}\,,
\end{split}\end{equation}
for the first two choices of coordinates, and
\begin{equation}\begin{split}\label{Eq:parameters3}
    a_0&=\frac{\mathfrak{s}}{2}+\frac{\ii (\omega-m\Omega_+)}{2 \kappa_+},\qquad a_1=\frac{\mathfrak{s}}{2}+\frac{\ii (\omega-m\Omega_-)}{2 \kappa_-}\,,
    \\a_t&=\frac{\mathfrak{s}}{2}+\frac{\ii (\omega-m\Omega_i)}{2 \kappa_i},\qquad a_\infty=\frac{\mathfrak{s}}{2}+\frac{\ii (\omega-m\Omega_h)}{2 \kappa_h}\,,
\end{split}\end{equation}
for the third choice. The surface gravity $\kappa_A$ and temperature $T_A$, at the singularity $A$, are defined as
\begin{equation}
    \kappa_A=\frac{\Delta_r'(R_A)}{2\left(1+\frac{\Lambda}{3}a_\text{BH}^2\right)\left(a_\text{BH}^2+R_A^2\right)},\qquad T_A=\frac{|\kappa_A|}{2\pi}\,.
\end{equation}
The surface gravity can be either positive or negative.

Finally, the accessory parameter reads\footnote{These expressions can be obtained from \autoref{tab:heun-crossing-accessory}, identifying the primed variables there $(u^\prime, t^\prime, a_0^\prime,\dots)$ with the unprimed ones in this section $(u,t,a_0,\dots)$, after attaching a subscript $``\text{there}"$ to the homonymous unprimed variables in \autoref{tab:heun-crossing-accessory}, e.g., $t\to t_\text{there}$, $u\to u_S =u_\text{there}$, etc.}
\begin{equation}\label{eq.usec4}
    u =u(t)=\begin{cases}&-\frac{1}{2}+2a_t^2-u_S\, \\ &\frac{t}{t-1}\,\biggl(-\frac{1}{2}+2a_t^2-u_S\biggr) \\ & \frac{(4 t-2) a_t^2-2 a_{\infty }^2-2 a_0^2+2 a_1^2-(t-1) \left(2 u_S+1\right)}{2 (t-1)}\end{cases},
\end{equation}
where the expression of $u_S:=u_\text{there}$ is given in \eqref{eq:UsC}. 
For later reference we note that all three choices of relations between our accessory parameter and the one of~\cite{Arnaudo:2025btb}, are such that in a large-$t$ expansion
\begin{equation}\label{eq:Asymputhree}
    u(t)=-\frac{1}{2}+2a_t^2-u_S+\mathcal{O}\left(\frac{1}{t}\right)\,.
\end{equation}
For a Kerr black hole, $u$ depends on angular momenta through the separation constant associated with the spin-weighted spheroidal harmonic \cite{Arnaudo:2025btb}
\begin{equation}
    {}_{\mathfrak{s}}A_{\ell m}={\ell(\ell+1)-\mathfrak{s}^{2}}-\frac{2m\left[\ell(\ell+1)+s^{2}\right]}{\ell(\ell+1)}\,a_\text{BH}\,\omega+\mathcal{O}\!\left(a_\text{BH}\sqrt{\Lambda}\right)\,,
\end{equation}
where $\ell \geq \text{max}(|\mathfrak{s}|,|m|)$ is the angular harmonic number and $\mathfrak{s}$ is the spin of the mode.

\paragraph{Fluctuations with Vanishing Frequency at Extremality.}
Let us look for solutions to the transcendental equations \eqref{eq:BasisDefinition} defining the spectrum of basis states $|\Phi_n\rangle$ in the near-cold expansion defined by
\begin{equation}
    R_h\to R_i,\qquad \Omega_h\to\Omega_i\to\Omega \qquad  T_{h},T_i\to0, \qquad t\to\infty\,,
\end{equation}
and
\begin{equation}
    \kappa_h =-\kappa_i\to 0^+,\qquad T_h\to 0\,.
\end{equation}
In this section, we will only be looking for wave-functions with low frequencies
\begin{equation}\label{eq:SmallFrequencyFinal}
    \omega =\widehat{\omega}{T}_h,\qquad\widehat{\omega}=\mathcal{O}(t^0)\,,
\end{equation}
at small temperatures (large $t$). 

For these modes, under the assumption $m\neq 0$, the critical exponents at the four singularities follow from \eqref{Eq:parameters}
\begin{equation}\label{eq:Lim1}\begin{split}
    a_0&\to\frac{\mathfrak{s}}{2}\pm m\frac{\ii (\Omega_\mp)}{4\pi T_\mp},\qquad \qquad\;\, a_1\to\frac{\mathfrak{s}}{2}\mp m\frac{\ii (\Omega_\pm)}{4\pi T_\pm}\,,\\
    a_t&\to\frac{\mathfrak{s}}{2}+m\frac{\ii (\Omega_h)}{4\pi T_h} \to \infty,\qquad a_\infty\to\frac{\mathfrak{s}}{2}-m\frac{\ii (\Omega_i)}{4\pi T_i}\to \infty\,,
\end{split}\end{equation}
and from~\eqref{Eq:parameters3}
\begin{equation}\label{eq:Lim2}\begin{split}
    a_0&\to\frac{\mathfrak{s}}{2}-m\frac{\ii (\Omega_+)}{4\pi T_+},\qquad \qquad\;\, a_1\to\frac{\mathfrak{s}}{2}+ m\frac{\ii (\Omega_-)}{4\pi T_\pm}\,,\\
    a_t&\to\frac{\mathfrak{s}}{2}-m\frac{\ii (\Omega_i)}{4\pi T_i} \to \infty,\qquad a_\infty\to\frac{\mathfrak{s}}{2}+m\frac{\ii (\Omega_h)}{4\pi T_h}\to \infty\,.
\end{split}\end{equation}
Since there is no dependence on $\widehat{\omega}$, such ansatz cannot give asymptotic solutions to \eqref{eq:BasisDefinitionTot} at large $t$. 
Thus, light modes can only exist for the choice of axial angular momentum
\begin{equation}\label{eq:Parameter2}
    m=0\,.
\end{equation}
In this case, we obtain
\begin{equation}\label{eq:P3}
    a_0\to\frac{\mathfrak{s}}{2},\qquad a_1\to\frac{\mathfrak{s}}{2},\qquad a_t\to\frac{\mathfrak{s}}{2}+\frac{\ii \widetilde{\omega}}{4\pi},\qquad a_\infty\to\frac{\mathfrak{s}}{2}-\frac{\ii \widetilde{\omega}}{4\pi}\,,
\end{equation}
where 
\begin{equation}\label{eq:SignDiff}
\widetilde{\omega}= \begin{cases}&+\widehat{\omega}\qquad \text{ for} \qquad \eqref{eq:Lim1}\\ &-\widehat{\omega} \qquad  \text{ for} \qquad \eqref{eq:Lim2}\end{cases}\,.
\end{equation}
Using this ansatz we evaluate the eigenvalue $\gamma^\text{out}$ of the outer bare differential operator $\widetilde{\mathcal{D}}_2^\text{out}$ \eqref{eq.tildeD2out} in the outer region, for all possible identifications~\eqref{eq:ChoicesRadialCoordinates}
\begin{equation}\begin{split}\label{eq:GammaFinal}
    \gamma^\text{out}=\lambda^\text{out}+\mathcal{O}\left(\frac{1}{t}\right)&=-a_t^2+a_\infty ^2-a_0^2-a_1^2+u+\frac{1}{2}+\mathcal{O}\left(\frac{1}{t}\right)\\
    &=+a_t^2+a_\infty^2-a_0^2-a_1^2-u_{S}+\mathcal{O}\left(\frac{1}{t}\right)\,\\
    &=-a_t^2+a_\infty^2-a_0^2-a_1^2+u_0+\frac{1}{2}\,,
\end{split}
\end{equation}
where the expression for $u_0$ can be found in~\eqref{eq:MatoneRel}. 

Clearly, at this point of the analysis, at leading order in the large-$t$ expansion, the three choices~\eqref{eq:ChoicesRadialCoordinates} are equivalent, up to the difference in sign choice in~\eqref{eq:SignDiff}. Expanding the expression of $u_S$ given in \eqref{eq:UsC}, using the second equality in~\eqref{eq:GammaFinal} and recalling the relation between $\kappa$ and $\gamma^\text{out}$ given in \eqref{eq:KappaGamma}, we find that the following relation holds for the light modes\footnote{This choice of $\pm$ is unrelated with the choices of radial coordinates~\eqref{eq:ChoicesRadialCoordinates}.}
\begin{equation}
    \kappa=\pm\frac{1}{2}\sqrt{(1+2\ell)^2}\,.
\end{equation}
As explained in \eqref{eq:Kappa}, whenever
\begin{equation}
    \text{Re}(\kappa)=-\left(\frac{1}{2}+\ell\right)<0\,,
\end{equation}
the only possible light eigenmodes defined by the transcendental equation \eqref{eq:BasisDefinition}, at small enough temperature $t\to \infty$, must have frequencies defined by one of the two conditions \eqref{eq:BasisDefinitionTot}
\begin{equation}\label{eq:BasisDefinitionTotFinal}
    \left(\pm a_\infty-\kappa+a_t+\frac{1}{2}\right)=-p, \qquad \text{or} \qquad \left(\pm a_{\infty}-\kappa-a_t+\frac{1}{2}\right)=-p,\qquad p=0,1,2,\dots\,.
\end{equation}
For the choice of $+$ ($-$) sign, only the second (first) condition is solvable. The solutions are
\begin{equation}
    2a_t=-k:=\mathfrak{s}\pm (\ell+p),\qquad p=0,1,2,\dots\,,
\end{equation}
or equivalently
\begin{equation}\begin{split}\label{eq:Frequencies}
    \frac{\ii\widetilde{\omega}}{2\pi}&= n := \pm \big(\ell+p\bigr)\,.
\end{split}
\end{equation}
Since $\ell \geq |\mathfrak{s}|$, the integer labels above satisfy the following inequalities for the $+$ sign
\begin{equation}\label{eq:AntiQNM}
    n\geq|\mathfrak{s}|\qquad\text{ and }\qquad k\geq\mathfrak{s}+|\mathfrak{s}|\,,
\end{equation}
and for the $-$ sign
\begin{equation}\label{eq:QNM}
    n\leq -|\mathfrak{s}|\qquad\text{ and }\qquad k\geq\mathfrak{s}-|\mathfrak{s}| \,.
\end{equation}
These solutions are consistent with the pseudo-square-integrability condition \eqref{eq:SqIntegrability} if and only if $\pm\text{Re}({a_{\infty}})>0$ , i.e., if and only if:
\begin{equation}\label{eq:SqIntFinSec}
    +:\quad \mathfrak{s}+n>0,\qquad -:\quad \mathfrak{s}-n>0\,,
\end{equation}
respectively for the two choices of sign. Finally,~\eqref{eq:SqIntFinSec} follows from \eqref{eq:AntiQNM} and \eqref{eq:QNM}, if and only if
\begin{equation}
    \mathfrak{s}\,\neq\, 0\,.
\end{equation}
This means that in $\mathcal{H}_\text{wzm}$ there are no states with $\mathfrak{s}=0$.

The two signs in \eqref{eq:Frequencies} can be interpreted as selecting QNM and anti-QNM frequencies in Lorentzian signature.\footnote{Note that the sign ambiguity in~\eqref{eq:SignDiff} interchanges these two subsets of modes.} On the other hand, in Euclidean signature, after Wick rotation and periodic identification of time $t_{L}\sim t_{L}+\frac{1}{\ii T_h}$, they are normal Matsubara modes. This follows from
\begin{equation}\label{eq:MatsubaraFreqs}
    \omega=-2\pi\ii T_h n+\text{subleading corrections},\qquad |n|\geq |\mathfrak{s}|,\quad  n,p\in\mathbb{Z}\,,
\end{equation}
where, as $2\kappa$ asymptotes to a positive integer at $T_h\to 0$, the subleading corrections may include only quadratic or higher powers of $T_h$, and logarithmic corrections in $T_h$ times polynomials in $T_h$ that vanish as $T_h\to 0$.

For $\mathfrak{s}=2$ ($\mathfrak{s}=1$) these pseudo-square-integrable light modes correspond to infinitely many copies of the tensor (vector) would-be-zero modes in the effective $AdS_2$ Jackiw--Teitelboim dual description of the Schwarzian theory \cite{Iliesiu:2022onk}.\footnote{Our analysis in this section is for Kerr solutions. The analysis in \cite{Iliesiu:2022onk} was for electrically charged but non-rotating solutions. Notice that rotation implies that the near-horizon geometry is a non-trivial fibration, not a direct product $AdS_2\times S^2$.} Different copies are labelled by the free integer index $p$ that runs from $0$ to $\infty$. At the level of the wave-function, this identification follows from using \eqref{eq:SeedHyper} and~\eqref{eq:SeedHyperInner}\footnote{Note that at $t\to\infty$ and fixed $\zeta$, the variable $z$ localizes to the horizon $z\to\infty$. In this sense we say that these $\mathrm{SL}(2,\mathbb{R})$-preserving modes are localized around the horizon. Of course, they will also have a non-trivial profile in the outer region which at large enough $t$ asymptotes to the hypergeometric function~\eqref{eq:SeedHyper}.}
\begin{equation}\label{eqLightwave-function}
    \Psi(z) = \begin{cases}\psi^\text{in}_\omega (\zeta)\,,& \quad \frac{z}{t} =\zeta \\ \psi^\text{out}_{\omega}(z) & \quad 0\leq z\ll t \end{cases} +\mathcal{O}\left(\frac{1}{t}\right)
\end{equation}
together with~\eqref{eq:P3},~\eqref{eq:SignDiff},~\eqref{eq:GammaFinal}, \eqref{eq:Frequencies} and finally~\eqref{eq:MatsubaraFreqs}. For $s=2$, the excluded modes $n=-1,0,1$ correspond to the generators of $\mathrm{SL}(2,\mathbb{R})$. For $s=1$, the excluded mode $n=0$ corresponds to the generator of $U(1)$.\footnote{These modes are precisely those giving the universal $\log T_h^{3/2}$ and $\log T_h^{1/2}$ corrections to thermal free energy, respectively. This follows from~\eqref{eq:MatsubaraFreqs}, which is equivalent to the Matsubara condition assumed in equation~(4.12) of~\cite{Arnaudo:2025btb}, as shown in the latter reference.}

\section{Discussion}\label{sec.5}
In this work we have proposed a method to compute near-extremal expansions of the spectrum and eigenfunctions of linear fluctuations around geometries for which separation of variables is possible. In order to illustrate its potential applications and to compare with other approaches, we have focused on radial problems for fluctuations described by a Heun differential equation. Concretely, we have shown that such problems project, via diffeomorphism actions, onto two isospectral Gauss hypergeometric problems in complementary regions of spacetime. The isospectral reduction fixes Robin boundary conditions at the intersection between the two complementary regions. Such boundary conditions are uniquely fixed by requiring the two diffeomorphisms to coincide in the matching region. 

The relevant Heun spectrum and eigenfunctions can be determined by considering the connection problem only in the inner (resp. outer) region, namely, in the near-horizon (resp. far) region of the geometry (e.g. black hole). All information regarding the outer (resp. inner) region is encoded in the Robin boundary conditions at the boundary of the near-horizon. The method reproduces subleading corrections which come from Heun connection coefficients, originally deduced using Nekrasov--Shatashvili expansions of correlators in two-dimensional Liouville field theory. As for the wave-functions, once the spectrum is solved, the wave-function of the original problem is recovered using the inverse action of the diffeomorphisms. In each of the two complementary regions, the answer reduces to an infinite sum over Gauss hypergeometric components. 

As an illustration of potential applications, we have used the method to identify explicitly a basis of pseudo-square-integrable low-frequency modes which spans the functional space of a Schwarzian theory in the low-temperature expansion of Kerr--de Sitter and anti-de Sitter black holes. Our notion of pseudo-square-integrability can be used to compute small-temperature expansions of gravitational path integrals of fluctuations around fixed gravitational geometries, such as black holes.

\paragraph{Future Directions.}
There are improvements, extensions and applications of the present work that we leave for the future. 
\begin{itemize}[parsep=0pt, listparindent=\parindent]
    \item The $\mathcal{O}(1/t)$ corrections to the light-mode wave-function~\eqref{eqLightwave-function} can be implemented as explained in \autoref{eq:Projection}. These corrections enforce the smooth transition between the inner and outer hypergeometric patches across the matching region in the radial variable domain, at fixed $t$. It would be important to numerically test such smooth transition. The analytic approach presented here implies that this has to be the case, but it would be reassuring to check it via explicit numerical computation. 
   
    \item For Kerr solutions, for example, we can search for other low-temperature modes with the aim of exploring potential quantum instabilities (e.g. superradiance) of the effective Schwarzian/JT description. We are currently investigating whether such modes exist and, if so, how their contribution to the thermodynamic free energy competes with that of the tensor and vector would-be zero modes in~\eqref{eqLightwave-function}.
    
    \item The approach presented in this paper can be applied to differential equations beyond the Heun equation, such as confluent Heun equations or equations with more than four regular singularities \cite{Denef:2000nb,Denef:2002ru,Bates:2003vx,Denef:2007vg,Arnaudo:2025kof}. These singularities may be irregular as well. For example, the idea behind our results applies also to a confluent Heun equation, whose interacting potential, in the outer region, contains terms of the form
    \begin{equation}
        Q_\text{int}^\text{out}(z)=\frac{\mu}{t z}+\frac{\nu}{t^2}+\dots\,.
    \end{equation}
    In the inner region the specifics of the computation are different, but the method still applies. In particular, the parameters $\mu$ and $\nu$ must be considered small perturbative parameters there, if one wishes to work with a hypergeometric equation at leading order. These perturbations can also be disconnected by a choice of diffeomorphism. 
    
    The same idea could be generalized to geometries where separation of variables is not yet fully achieved, such as Kerr-Newman solutions. In those cases it seems natural to explore the use of interaction-disconnecting diffeomorphisms whose radial and angular dependence is not separable.
    
    \item It would be interesting to compare the perspective developed here on the gravitational path integral over black hole fluctuations in de Sitter spacetimes, which is agnostic about the choice of time signature, with the perspectives of \cite{Ivo:2025yek,Turiaci:2025xwi}, which deal with gravitational path integrals of fluctuations about de Sitter.\footnote{The numerical explorations reported in section 3.2 of \cite{Turiaci:2025xwi} are consistent with our analytic conclusions. See, for instance, \autoref{sec.4}.} The computation of path integrals of fluctuations around de Sitter should be recoverable from the black hole computation in limits where all physical horizons become small with respect to the cosmological one. 
  
    \item It should be possible to formulate our construction directly in the language of Liouville CFT and the AGT correspondence \cite{Alday:2009aq}. In the semiclassical limit, the Heun equation arises as the BPZ equation for a chiral five-point block with a level-two degenerate insertion, while the two hypergeometric problems correspond to four-point degenerate blocks on the components of the degenerate sphere. Our result suggests that the full five-point block may be computed within a single component, with the contribution of the complementary one encoded in Robin boundary conditions. Frobenius boundary conditions select definite degenerate fusion channels, whereas the Robin data fix a modulus-dependent linear combination of them across the plumbing neck. Through AGT, this should translate into the Nekrasov--Shatashvili limit of the $SU(2)$ theory with $N_f=4$ in the presence of a surface defect \cite{Alday:2009fs}: the full defect partition function could be reconstructed from the hypergeometric defect partition function of a theory without a dynamical four-dimensional gauge group, with the effects of the $SU(2)$ gauging and of the complementary matter sector encoded in the Robin coefficient. 

    \item Finally, we could apply our Heun-to-hypergeometric projection to the spectral problem on toric Sasaki--Einstein spaces. In particular, the two coupled Heun equations governing scalar harmonics on the $L^{a,b,c}$ manifolds could be replaced by hypergeometric problems defined on complementary patches of the corresponding physical intervals. Regularity at the degeneration loci of the torus fibration would select the admissible local solutions, while Robin boundary conditions at the interfaces would encode the effects of the complementary patches. Requiring the two projected problems to be simultaneously isospectral could then determine both the Laplacian eigenvalue and the common separation constant. The $Y^{p,q}$ limit, in which one of the Heun equations reduces exactly to a Gauss hypergeometric equation, provides a controlled starting point for developing this construction perturbatively away from the cohomogeneity-one locus. Such an approach could provide analytic access to generic non-polynomial Kaluza--Klein modes and, through the AdS/CFT correspondence, to the conformal dimensions of non-BPS operators in the dual quiver gauge theories.

\end{itemize}

\paragraph*{Acknowledgments.}
We thank Paolo Arnaudo, Cristoforo Iossa and Giulio Bonelli for useful discussions. We thank the Galileo Galilei Institute for Theoretical Physics for hospitality during the workshop “Pathways to Quantum Black Holes: from Effective Theories to Exact Methods”. The work of ACB was supported by the CRT Foundation under the grant 108399/2024.0434, and by the Munich Institute for Astro-, Particle and BioPhysics (MIAPbP) which is funded by the Deutsche Forschungsgemeinschaft (DFG, German Research Foundation) under Germany's Excellence Strategy – EXC-2094 – 390783311. ACB, DM and LR acknowledge partial support by the INFN.

\appendix

\section{Transformations of the Heun Equation}\label{app:Relabels}
An analogous discussion applies to a small-$t$ expansion of $\mathcal{L}_2$ \eqref{eq.L2}. In this case, the fixed $z$ expansion selects the inner region centred around $z=\infty$. Namely, the expansion $t\to 0$ does commute with the $z\to \infty$ expansion, but it does not commute with the $z\to 0$ expansion. To cover the dual region concentrated around $z=0$, a natural radial coordinate is $t\,\zeta=z$, which preserves the critical exponents at $z=\zeta=0$ but changes the exponent at $z=\infty$ to $\zeta=\infty$. We have checked that this description can be mapped to the previous one, using $\mathrm{SL}(2,\mathbb{Z})$ transformations such as the one summarized in the table below. Thus, without loss of generality in the main body of the paper we focused on the ``$t\to\infty$'' description.
\begin{table}[ht]
\centering
    \small
    \renewcommand{\arraystretch}{1.35}
    \begin{tabular}{c|c|c|c}
    \hline
    Transformation & Puncture exchange & $t'$ & Parameter map \\
    \hline
    $z\mapsto z^\prime=1/z$ & $0\leftrightarrow \infty$ & $\dfrac{1}{t}$ & $ \begin{aligned}
        a_0'&=a_\infty, & a_1'&=a_1,\\
        a_{t'}'&=a_t, & a_\infty'&=a_0
    \end{aligned}$
    \\
    \hline
    $z\mapsto z^{\prime}=\dfrac{z-1}{z}$ & $\{0,1,t,\infty\}\leftrightarrow \{\infty,0,t^\prime,1\}$ & $\dfrac{t-1}{t}$ & $\begin{aligned} a_0'&=a_1, & a_1'&=a_\infty,\\
    a_{t'}'&=a_t, & a_\infty'&=a_0
    \end{aligned}$
    \\
    \hline
     $z\mapsto z^{\prime}=\dfrac{z-1}{z-t}$ & $\{0,1,t,\infty\}\leftrightarrow \{t^\prime,0,\infty,1\}$ & $\dfrac{1}{t}$ & $\begin{aligned} a_0'&=a_1, & a_1'&=a_\infty,\\
    a_{t'}'&=a_0, & a_\infty'&=a_t
    \end{aligned}$
    \\
    \hline
    \end{tabular}
    \caption{Basic crossing transformation of the Schrödinger-form Heun equation. These were computed using \eqref{eq:NonLinearDiff0}.}
    \label{tab:heun-crossing-basic}
\end{table}

\begin{table}[ht]
    \centering
    \small
    \renewcommand{\arraystretch}{1.55}
    \begin{tabular}{c|c|c}
    \hline
    Transformation & $t'$ & $u'$ \\
    \hline
    $z\mapsto z^\prime= 1/z$ & $\displaystyle \frac{1}{t}$ & $\displaystyle -\frac{1}{2} + 2a_t^2 - u$ \\
    \hline
     $z\mapsto z^\prime= \frac{z-1}{z}$ & $\displaystyle \frac{t-1}{t}$ & $\displaystyle -\frac{1}{2} (t-1) \left(4 a_t^2-2 u-1\right)=\frac{\left(-4 a_t^2+2 u+1\right) t'}{2-2 t'}$ \\
    \hline
     $z\mapsto z^\prime= \frac{z-1}{z-t}$ & $\displaystyle \frac{1}{t}$ & $\displaystyle \frac{a_0^2 \left(4 t'-2\right)-2 a_t^2+2 a_{\infty }^2-2 a_1^2-(2 u+1)
   \left(t'-1\right)}{2 \left(t'-1\right)}$ \\
    \hline
    \end{tabular}
    \caption{Transformation of the Heun modulus $t$ and accessory parameter $u$.}
    \label{tab:heun-crossing-accessory}
\end{table}

Using the Schrödinger transformation law \eqref{eq:NonLinearDiff0} with the change of variable
\begin{equation}
    z=F[z^\prime]=\frac{1}{z^\prime}\,,
\end{equation}
we obtain
\begin{equation}\label{eq:TransformationPotentialApp}
    Q_\text{dual}\left(z^\prime\right)=\frac{1}{z^{\prime 4}}Q\left(\frac{1}{z^\prime};a_0,a_1,a_t,a_\infty,u,t\right)\,,
\end{equation}
where $Q$ is given by \eqref{eq.schroedinger}. We then notice that
\begin{equation}
    Q_\text{dual}(z^\prime) =  Q(z^\prime,a_{\infty},a_1,a_t,a_0,u^\prime,t^\prime)\,.
\end{equation}
Thus, in summary we have the ``self-duality'' identity 
\begin{equation}\label{eq:SelfDuality}
    Q(z^\prime,a_{0}^\prime,a_1^\prime,a_t^\prime,a_0^\prime,u^\prime,t^\prime)=\frac{1}{z^{\prime 4}}Q\!\left(z=\frac{1}{z^\prime};a_0,a_1,a_t,a_\infty,u,t=\frac{1}{t^\prime}\right)\,.
\end{equation}
The primed variables are defined in terms of the unprimed ones in \autoref{tab:heun-crossing-basic} and \autoref{tab:heun-crossing-accessory}.

Suppose we apply the isospectral reduction to the hypergeometric equations obtained by expanding $t^\prime=1/t\to \infty$ of $Q(z^\prime,a_{0}^\prime,a_1^\prime,a_t^\prime,a_0^\prime,u^\prime,t^\prime)$. Then the problem is equivalent to the one developed in the main text for the expansion $t\to \infty$ of $Q(z,a_{0},a_1,a_t,a_0,u,t)$. Suppose the eigenvalues found in that approach have the form
\begin{equation}\label{eq:Problem1}
\lambda_n=\Lambda_n(a_0,a_1,a_t,a_0,u,t)\,,
\end{equation}
then \eqref{eq:SelfDuality} implies that the eigenvalues of the primed problem are functionally related to the eigenvalues of the unprimed problem as follows
\begin{equation}
\lambda^\prime_n=\Lambda_n(a_0^\prime,a_1^\prime,a_t^\prime,a_0^\prime,u^\prime,t^\prime)\,,
\end{equation}
where $\Lambda_n$ is the same function of six variables as in~\eqref{eq:Problem1}.

\section{On Uniqueness of the Diffeomorphism}
Solutions to the homogeneous part of equation \eqref{eq:DiffEqFlucConc} can be found in terms of solutions to the hypergeometric Schrödinger equation
\begin{equation}
    \left[\partial_Z^2+Q_0(Z)\right]\chi(Z)=0 .
\end{equation}
Indeed, if $\chi_1,\chi_2$ solve the latter, then
\begin{equation}
    \chi_1^2\,,\qquad\chi_1\chi_2,\qquad\chi_2^2
\end{equation}
solve the third-order homogeneous version of \eqref{eq:DiffEqFlucConc}. Let us use the basis of hypergeometric solutions that is Frobenius at $Z=0$
\begin{equation}
    \phi_+^\text{out}(Z)=Z^{\frac{1}{2}+a_0}(1-Z)^{\frac{1}{2}+a_1}{}_2F_1\left(\frac{1}{2}+a_0+a_1+\kappa,\frac{1}{2}+a_0+a_1-\kappa;1+2a_0;Z\right)\,,
\end{equation}
\begin{equation}
    \phi_-^\text{out}(Z)=Z^{\frac{1}{2}-a_0}(1-Z)^{\frac{1}{2}+a_1}{}_2F_1\left(\frac{1}{2}-a_0+a_1+\kappa,\frac{1}{2}-a_0+a_1-\kappa;1-2a_0;Z\right),
\end{equation}
where we recall
\begin{equation}
    \kappa^2=a_\infty^2-a_t^2+u+\frac{1}{4}\,.
\end{equation}
Then the homogeneous solutions may be written as a linear combination of the three independent solutions
\begin{equation}
    h_+^{(0)}=(\phi_+^\text{out})^2,\qquad h_0^{(0)}=\phi_+^\text{out}\phi_-^\text{out},\qquad h_-^{(0)}=(\phi_-^\text{out})^2\,.
\end{equation}
Thus, the most general solution to the inhomogeneous equation is
\begin{equation}
    f_1^\text{out}(Z)=f_\text{p}^\text{out}(Z)+B_+(\phi_+^\text{out})^2+B_0\phi_+^\text{out}\phi_-^\text{out}+B_-(\phi_-^\text{out})^2\,.
\end{equation}
From regularity at $Z=0$, and assuming $a_0>0$, we find
\begin{equation}\label{eq:BZero}
    B_-=0\,.
\end{equation}
From demanding leading behaviour of the form $\sim Z$ at $Z=\infty$, and assuming $\kappa>0$, we find 
\begin{equation}\label{eq:Bplus}
    B_+=-\frac{B_0\,\mathcal{C}_{-+}^{(\infty,0)}}{\mathcal{C}_{++}^{(\infty,0)}}\,.
\end{equation}
This follows from using the connection formula for $\phi_\pm^\text{out}$ with the Frobenius solutions at $Z=\infty$. The label $\pm$ in $\mathcal{C}_{\pm+}^{(\infty,0)}$ denotes the connection coefficient between $\phi_\pm^\text{out}$ and the Frobenius solution at $Z=\infty$ with critical exponents $\frac{1}{2}\pm\kappa$ there, with canonical normalization.

This implies that, if one wants only analytic behaviour of the homogeneous solution in a perturbative expansion around $Z=0$, as well as $Z=\infty$, it is necessary to fix the trivial choice
\begin{equation}\label{eq:TrivialBs}
    B_+=B_-=B_0=0\,.
\end{equation}
Around $Z=0$, this is because $\phi_+^\text{out}$ has non-integer powers of the form $Z^{n+2a_0}$, with $n$ positive integers. Around $Z=\infty$ a non-trivial homogeneous solution respecting \eqref{eq:BZero}-\eqref{eq:Bplus} will have powers of the form $Z^{-n-\kappa}$ unless \eqref{eq:TrivialBs} is satisfied.

\section{Interactions via Green-Function Analysis}\label{app:GreenF}
If we do not work with the coordinates $Z$ and $\wp$ in the outer and inner regions, we need to compute radiative corrections. In this appendix, we show how to compute such corrections using twisted Green functions. 

\subsection{Born-Series Construction of Perturbed Solutions}
Let the perturbed operator, which may be non-Hermitian, be split into leading and subleading interacting contributions
\begin{equation}
    \mathcal{D}_2 =\D_2^{(0)}-V_\text{int}\,,
\end{equation}
where $V_\text{int}=\mathcal{O}(\varepsilon)$ with $\varepsilon\to0$ as $t\to \infty$. We seek a solution of
\begin{equation}
    \D_2\Psi = 0
\end{equation}
in the form
\begin{equation}
    \Psi = \sum_{\ell=0}^\infty \varepsilon^\ell\psi^{(\ell)} \,,
\end{equation}
where we define the coefficient functions $\psi^{(\ell)}$ iteratively, starting from
\begin{equation}
    \D^{(0)}_2 \psi^{(0)} = 0.
\end{equation}
and, at each integer level, equate $\ell\geq 1$
\begin{equation}
    \varepsilon \D_2^{(0)} \psi^{(\ell+1)} = V_\text{int}\psi^{(\ell)}\,.
\end{equation}
Upon resummation, we obtain the desired formula
\begin{equation}\label{eq:BornSeries}
    |\Psi\rangle=\sum_{\ell=0}^{\infty}\biggl({{\widehat{\mathcal{D}}^{(0),-1}_2 \widehat{V}_\text{int}}}\biggr)^\ell |\psi^{(0)}\rangle=\frac{1}{\mathbb{I}- {\widehat{\mathcal{D}}^{(0),-1}_2 \widehat{V}_\text{int}} }\,|\psi^{(0)}\rangle
\end{equation}
where the inverse operator was defined in equation \eqref{eq:InverseOp}. Notice that in this formula there is no explicit dependence on the parameter $\varepsilon$. We have left the hats in this equation to highlight the fact that these are operators acting on kets. The geometric series resummation in~\eqref{eq:BornSeries} is understood in operatorial sense.

We choose the initial condition
\begin{equation}
    \psi^{(0)} = \chi(z)\equiv \langle z|\chi\rangle\,.
\end{equation}
The perturbed solution $\Psi$ receives dynamical corrections in the large-$t$ expansion coming from $V_\text{int}$:
\begin{equation}
    \Psi=\psi^{(0)}+\frac{1}{t}\psi^{(1)}(z)+\dots\,.
\end{equation}
To illustrate, we expand \eqref{eq:BornSeries} at first order
\begin{equation}\
    \psi^{(1)}(z)=\langle z|\mathcal{D}_2^{-1} V_\text{int}|\chi\rangle=\sum_{n\geq 0}\phi_{n}(z)\,\frac{\langle\overline{\phi}_{n}|\,V_\text{int}\,|\chi\rangle}{\gamma_{n}-\lambda} +\mathcal{O}\left(\frac{1}{t}\right)\,,
\end{equation}
where
\begin{equation}
\langle\overline{\phi}_{n}|\,V_\text{int}\,|\chi\rangle\equiv\int_\omega \, \overline{\phi}_n(z) V_\text{int}(z)\chi(z)\,\dd z\,.
\end{equation}
These integrals involve hypergeometric functions and a rational function $V_\text{int}$. They can be evaluated numerically, in a perturbative expansion in $1/t$, or using truncations at sufficiently large $t$. By construction, we assume that the interaction term does not make these integrals diverge. The integration domains are given by
\begin{equation}
    \Gamma=
    \begin{cases}
        \quad (0,1)\,, & \text{ outer region},\\
        \quad (1,\infty)\,, & \text{ inner region}.
\end{cases}
\end{equation}
We do not need to assume that the hypergeometric initial wave-function $\chi$ is a single element of the twisted discrete basis; it may instead be a more complicated wave-function, for example an infinite linear combination of basis elements. The choice depends on the physical problem of interest.

\subsection{Perturbative Expansion}
If $V_\text{int}$ depends on the eigenvalue $\lambda$, which is what happens in black hole examples, the spectral problem becomes non-linear. In that case the discrete spectrum for $\lambda$ is determined by
\begin{equation}
    \det{\left(\mathbb{H}_{mn}(\lambda)-\lambda\delta_{mn}\right)}=0\,.
\end{equation}
 We define the matrix representation of $\mathbb{H}_{mn}$ of $\mathcal{D}_2$ in the twisted basis as follows
\begin{equation}
    \mathbb{H}_{mn}=\gamma_n\,\delta_{mn}-\langle\overline{\phi}_n|V_\text{int}|\phi_m\rangle\,.
\end{equation}
The $1/t$ perturbative expansion linearizes this problem by using the ansätze
\begin{equation}
    \lambda=\sum_{\ell=0}^{\infty}\frac{\lambda^{(\ell)}}{t^\ell}, \qquad \lambda^{(0)}=\gamma_n\,\qquad n=0,\dots,\infty\,.
\end{equation}
Even in this expansion, it is easy to verify that this way of computing the spectrum is not computationally efficient. 

The Born-series approach simplifies the analysis. Expanding \eqref{eq:BornSeries} with initial condition given by a twisted basis element $\psi^{(0)}=\phi_n$, and projecting it onto the twisted basis, we obtain the expansion
\begin{equation}\label{eq:ExpansionhypergeometricBasis}
    \Psi_\lambda = \sum_{m=0}^{\infty} c_m(\lambda) \phi_m\,,
\end{equation}
for a wave-function in the kernel of $\mathcal{D}_2-\lambda$
 \begin{equation}
     (\mathcal{D}_2-\lambda)\,\Psi_{\lambda}=0\,.
 \end{equation}
The 
form of $c_m=c_m(\lambda)$ follows from \eqref{eq:BornSeries}, the expression for $V_\text{int}$, and the twisted Green function
\begin{equation}\label{eq:Cm}
    c_m=\delta_{mn}+\frac{V_{mn}}{\lambda-\gamma_m}+\sum_{k=0}^{\infty}\frac{V_{mk}V_{kn}}{(\lambda-\gamma_m)(\lambda-\gamma_k)}+\sum_{k=0}^{\infty}\sum_{\ell=0}^{\infty}\frac{V_{mk}V_{k\ell}V_{\ell n}}{(\lambda-\gamma_m)(\lambda-\gamma_k)(\lambda-\gamma_\ell)}+\mathcal{O}(V^4)\,,
\end{equation}
where $V_{mn}:=-\langle \overline{\phi}_m|V_\text{int}|\phi_n\rangle$.

The eigenvalues of these functions are related to the eigenvalues of the selected twisted basis element by the following integral and non-linear relations (following from~\eqref{eq:Cm}), which are expected to be related to the integrable structures pointed out in \cite{fioravanti2021newmethodexactresults}:
\begin{equation}\label{eq:EigenvaluesSource}
    (c_m-\delta_{mn}) (\lambda-\gamma_m)-\sum_{k=0}^{\infty}V_{mk}c_k=0\,.
\end{equation}
This applies to eigenvalues of eigenfunctions $\Psi_{\lambda}$ living in the space of pseudo-square-integrable functions with respect to the bilinear product~\eqref{eq:Measure} and for potentials such that $V_\text{int} \phi_m(z)$ belongs to such a space as well
\begin{equation}
    V_\text{int}\, \phi_m(z) =- \sum_{k=0}^\infty V_{mk}\, \phi_{k}(z)\,.
\end{equation}
Finally, using~\eqref{eq:Cm} in the relation~\eqref{eq:EigenvaluesSource}, for $m=n$, (after some algebra that we do not report in here) gives the condition for eigenvalues $\lambda$
\begin{equation}
    \lambda=\gamma_n+V_{nn}+\sum_{k\neq n}\frac{V_{n k}V_{kn}}{\lambda-\gamma_k}+\sum_{\substack{k\neq n\\ \ell\neq n}}\frac{V_{nk}V_{k\ell}V_{\ell n}}{(\lambda-\gamma_k)(\lambda-\gamma_\ell)}+\dots\,.
\end{equation}
The isospectral deformation to hypergeometric problems bypasses these complications as it projects all matrix elements $V_{mn}$ to $0$, and thus, it corresponds to a choice of twisted basis where interactions vanish. In the analysis in this section we have assumed no implicit dependence on $t$ in parameters such as $u$, so this would imply $\lambda=\gamma_n$ once interactions are disconnected. 

The analysis assuming isomonodromic dependence on $t$ in $u=u(t)$, $\lambda_\text{iso}$, is related to the analysis without $t$-dependence in this section, $\lambda_\text{here}$, by the following redefinitions
\begin{equation}\label{eq:RedefLambda}
\begin{split}
\lambda_\text{here}&\,\equiv\, \lambda^\text{out}_{\text{iso}}-u(t)+u_0\,, \\
\lambda _\text{here}&\,\equiv \lambda^\text{in}_{\text{iso}}+u(t)-u_0\,,
\end{split}
\end{equation}
depending on whether one works in the outer or inner region. This follows because the right-hand sides are, by definition, independent of $t$, and the corresponding redefinitions become trivial as $t\to\infty$. Thus, from~\eqref{eq:RedefLambda} it follows that the answer for $\lambda$ in the case of an isomonodromic deformation, for a choice of twisted basis where all perturbative interactions have been disconnected $V_{nm}=0$, is
\begin{equation}\begin{split}
    \lambda^\text{out}-u&=\gamma^\text{out}-u_0\,,  \\
    \lambda^\text{in}+u&=\gamma^\text{in}+u_0\,.
\end{split}\end{equation}

\section{Kerr-(A)dS Accessory Parameter}
\begin{equation}\begin{split}\label{eq:UsC}
    u_{S}=-&\frac{\mathfrak{s}\left(\Lambda a_{\text{BH}}^2-3\right)}{\Lambda\left(R_h-R_+\right)\left(R_h+2R_i+R_+\right)}+\frac{3\, {}_{\mathfrak{s}}A_{lm}}{\Lambda\left(R_h-R_+\right)\left(R_i-R_-\right)}-\\
    &-\frac{1}{2\Lambda^2\left(R_h-R_-\right)\left(R_h-R_+\right)^3\left(R_i-R_-\right)\left(R_h-R_i\right)^2}\Bigg[\Lambda^2\Big(4a_{\text{BH}}^6\big(2\omega^2 R_h\left(R_h^2-R_+R_i\right)+m^2\left(3R_h+R_-\right)\big)\\
    &\qquad+4\ii m\mathfrak{s}\,a_{\text{BH}}^3\left(R_h-R_+\right)^2\left(R_h-R_i\right)^2-8m\omega a_{\text{BH}}^5R_h\left(R_h^2-R_+R_i\right)\\
    &\qquad-4\omega a_{\text{BH}}^4\Big(\omega R_h^3\big(2R_+R_i-R_h\left(R_i+R_+\right)\big)+\ii\mathfrak{s}\left(R_h-R_+\right)^2\left(R_h-R_i\right)^2\Big)\\
    &\qquad-4\mathfrak{s} a_{\text{BH}}^2\left(R_h-R_+\right)^2\left(R_h-R_i\right)^2\left(R_h\left(2+iR_-\omega\right)+R_i+R_+\right)\\
    &\qquad-8m\omega a_{\text{BH}}^7\left(3R_h+R_-\right)+4\omega^2 a_{\text{BH}}^8\left(3R_h+R_-\right)\\
    &\qquad+(\mathfrak{s}+1)\left(R_h-R_-\right)\left(R_h-R_+\right)^2\left(R_h-R_i\right)^2\\
    &\qquad\qquad\times\Big(-\mathfrak{s}\left(-2R_-R_h+R_i^2+R_+^2\right)-\left(R_i+R_+\right)^2\Big)\Big)\\
    &\qquad+6\Lambda\Big(4a_{\text{BH}}^4\big(2\omega^2R_h\left(R_h^2-R_+R_i\right)+m^2\left(3R_h+R_-\right)\big)\\
    &\qquad+2\ii m\mathfrak{s}\,a_{\text{BH}}\left(R_h-R_+\right)^2\left(R_h-R_i\right)^2-8m\omega a_{\text{BH}}^3R_h\left(R_h^2-R_+R_i\right)\\
    &\qquad+2\omega a_{\text{BH}}^2\Big(2\omega R_h^3\big(R_h\left(R_i+R_+\right)-2R_+R_i\big)-\ii\mathfrak{s}\left(R_h-R_+\right)^2\left(R_h-R_i\right)^2\Big)\\
    &\qquad-8m\omega a_{\text{BH}}^5\left(3R_h+R_-\right)+4\omega^2a_{\text{BH}}^6\left(3R_h+R_-\right)\\
    &\qquad+2\mathfrak{s}\left(R_h-R_+\right)^2\left(R_h-R_i\right)^2\left(R_h\left(2-\ii R_-\omega\right)+R_i+R_+\right)\Big)\\
        &\qquad+36\left(ma_{\text{BH}}-\omega\left(a_{\text{BH}}^2+R_h^2\right)\right)\\
    &\qquad\qquad\times\left(ma_{\text{BH}}\left(3R_h+R_-\right)-\omega a_{\text{BH}}^2\left(3R_h+R_-\right)-\omega R_h\big(R_h\left(R_i+R_+\right)-2R_+R_i\big)\right)\Bigg]\,.
\end{split}\end{equation}

\bibliographystyle{utphys}

\begin{thebibliography}{10}

\bibitem{Strominger:1998yg}
A.~Strominger, ``{AdS(2) quantum gravity and string theory},'' \href{http://dx.doi.org/10.1088/1126-6708/1999/01/007}{{\em JHEP} {\bfseries 01} (1999) 007}, \href{http://arxiv.org/abs/hep-th/9809027}{{\ttfamily arXiv:hep-th/9809027}}.

\bibitem{MaldacenaMichelsonStrominger1999}
J.~M. Maldacena, J.~Michelson, and A.~Strominger, ``Anti-de sitter fragmentation,'' \href{http://dx.doi.org/10.1088/1126-6708/1999/02/011}{{\em Journal of High Energy Physics} {\bfseries 1999} no.~02, (1999) 011}, \href{http://arxiv.org/abs/hep-th/9812073}{{\ttfamily arXiv:hep-th/9812073 [hep-th]}}.

\bibitem{BardeenHorowitz1999}
J.~M. Bardeen and G.~T. Horowitz, ``The extreme kerr throat geometry: A vacuum analog of \(ads_2 \times s^2\),'' \href{http://dx.doi.org/10.1103/PhysRevD.60.104030}{{\em Physical Review D} {\bfseries 60} (1999) 104030}, \href{http://arxiv.org/abs/hep-th/9905099}{{\ttfamily arXiv:hep-th/9905099 [hep-th]}}.

\bibitem{Sen:2005wa}
A.~Sen, ``{Black hole entropy function and the attractor mechanism in higher derivative gravity},'' \href{http://dx.doi.org/10.1088/1126-6708/2005/09/038}{{\em JHEP} {\bfseries 09} (2005) 038}, \href{http://arxiv.org/abs/hep-th/0506177}{{\ttfamily arXiv:hep-th/0506177}}.

\bibitem{Sen:2007qy}
A.~Sen, ``{Black Hole Entropy Function, Attractors and Precision Counting of Microstates},'' \href{http://dx.doi.org/10.1007/s10714-008-0626-4}{{\em Gen. Rel. Grav.} {\bfseries 40} (2008) 2249--2431}, \href{http://arxiv.org/abs/0708.1270}{{\ttfamily arXiv:0708.1270 [hep-th]}}.

\bibitem{Kunduri:2013gce}
H.~K. Kunduri and J.~Lucietti, ``{Classification of near-horizon geometries of extremal black holes},'' \href{http://dx.doi.org/10.12942/lrr-2013-8}{{\em Living Rev. Rel.} {\bfseries 16} (2013) 8}, \href{http://arxiv.org/abs/1306.2517}{{\ttfamily arXiv:1306.2517 [hep-th]}}.

\bibitem{Sen:2008yk}
A.~Sen, ``{Entropy Function and AdS(2) / CFT(1) Correspondence},'' \href{http://dx.doi.org/10.1088/1126-6708/2008/11/075}{{\em JHEP} {\bfseries 11} (2008) 075}, \href{http://arxiv.org/abs/0805.0095}{{\ttfamily arXiv:0805.0095 [hep-th]}}.

\bibitem{Sen:2008vm}
A.~Sen, ``{Quantum Entropy Function from AdS(2)/CFT(1) Correspondence},'' \href{http://dx.doi.org/10.1142/S0217751X09045893}{{\em Int. J. Mod. Phys. A} {\bfseries 24} (2009) 4225--4244}, \href{http://arxiv.org/abs/0809.3304}{{\ttfamily arXiv:0809.3304 [hep-th]}}.

\bibitem{Almheiri:2014cka}
A.~Almheiri and J.~Polchinski, ``{Models of AdS$_2$ backreaction and holography},'' \href{http://dx.doi.org/10.1007/JHEP11(2015)014}{{\em JHEP} {\bfseries 11} (2015) 014}, \href{http://arxiv.org/abs/1402.6334}{{\ttfamily arXiv:1402.6334 [hep-th]}}.

\bibitem{Jensen:2016pah}
K.~Jensen, ``{Chaos in AdS$_2$ Holography},'' \href{http://dx.doi.org/10.1103/PhysRevLett.117.111601}{{\em Phys. Rev. Lett.} {\bfseries 117} no.~11, (2016) 111601}, \href{http://arxiv.org/abs/1605.06098}{{\ttfamily arXiv:1605.06098 [hep-th]}}.

\bibitem{Maldacena:2016upp}
J.~Maldacena, D.~Stanford, and Z.~Yang, ``{Conformal symmetry and its breaking in two dimensional Nearly Anti-de-Sitter space},'' \href{http://dx.doi.org/10.1093/ptep/ptw124}{{\em PTEP} {\bfseries 2016} no.~12, (2016) 12C104}, \href{http://arxiv.org/abs/1606.01857}{{\ttfamily arXiv:1606.01857 [hep-th]}}.

\bibitem{Engelsoy:2016xyb}
J.~Engels{\"o}y, T.~G. Mertens, and H.~Verlinde, ``{An investigation of AdS$_2$ backreaction and holography},'' \href{http://dx.doi.org/10.1007/JHEP07(2016)139}{{\em JHEP} {\bfseries 07} (2016) 139}, \href{http://arxiv.org/abs/1606.03438}{{\ttfamily arXiv:1606.03438 [hep-th]}}.

\bibitem{Iliesiu:2020qvm}
L.~V. Iliesiu and G.~J. Turiaci, ``{The statistical mechanics of near-extremal black holes},'' \href{http://dx.doi.org/10.1007/JHEP05(2021)145}{{\em JHEP} {\bfseries 05} (2021) 145}, \href{http://arxiv.org/abs/2003.02860}{{\ttfamily arXiv:2003.02860 [hep-th]}}.

\bibitem{Iliesiu:2022onk}
L.~V. Iliesiu, S.~Murthy, and G.~J. Turiaci, ``{Revisiting the logarithmic corrections to the black hole entropy},'' \href{http://dx.doi.org/10.1007/JHEP07(2025)058}{{\em JHEP} {\bfseries 07} (2025) 058}, \href{http://arxiv.org/abs/2209.13608}{{\ttfamily arXiv:2209.13608 [hep-th]}}.

\bibitem{Faulkner:2009wj}
T.~Faulkner, H.~Liu, J.~McGreevy, and D.~Vegh, ``{Emergent quantum criticality, Fermi surfaces, and AdS(2)},'' \href{http://dx.doi.org/10.1103/PhysRevD.83.125002}{{\em Phys. Rev. D} {\bfseries 83} (2011) 125002}, \href{http://arxiv.org/abs/0907.2694}{{\ttfamily arXiv:0907.2694 [hep-th]}}.

\bibitem{Ren:2012hg}
J.~Ren, ``{Analytic quantum critical points from holography},'' \href{http://arxiv.org/abs/1210.2722}{{\ttfamily arXiv:1210.2722 [hep-th]}}.

\bibitem{Maldacena:1997re}
J.~M. Maldacena, ``{The Large $N$ limit of superconformal field theories and supergravity},'' \href{http://dx.doi.org/10.4310/ATMP.1998.v2.n2.a1}{{\em Adv. Theor. Math. Phys.} {\bfseries 2} (1998) 231--252}, \href{http://arxiv.org/abs/hep-th/9711200}{{\ttfamily arXiv:hep-th/9711200}}.

\bibitem{Fidkowski:2003nf}
L.~Fidkowski, V.~Hubeny, M.~Kleban, and S.~Shenker, ``{The Black Hole Singularity in AdS/CFT},'' \href{http://dx.doi.org/10.1088/1126-6708/2004/02/014}{{\em JHEP} {\bfseries 02} (2004) 014}, \href{http://arxiv.org/abs/hep-th/0306170}{{\ttfamily arXiv:hep-th/0306170 [hep-th]}}.

\bibitem{Festuccia:2005pi}
G.~Festuccia and H.~Liu, ``{Excursions beyond the Horizon: Black Hole Singularities in Yang--Mills Theories. I},'' \href{http://dx.doi.org/10.1088/1126-6708/2006/04/044}{{\em JHEP} {\bfseries 04} (2006) 044}, \href{http://arxiv.org/abs/hep-th/0506202}{{\ttfamily arXiv:hep-th/0506202 [hep-th]}}.

\bibitem{Dodelson:2023vrw}
M.~Dodelson, C.~Iossa, R.~Karlsson, and A.~Zhiboedov, ``{A thermal product formula},'' \href{http://dx.doi.org/10.1007/JHEP01(2024)036}{{\em JHEP} {\bfseries 01} (2024) 036}, \href{http://arxiv.org/abs/2304.12339}{{\ttfamily arXiv:2304.12339 [hep-th]}}.

\bibitem{Dodelson:2025jff}
M.~Dodelson, C.~Iossa, and R.~Karlsson, ``{Bouncing off a stringy singularity},'' \href{http://arxiv.org/abs/2511.09616}{{\ttfamily arXiv:2511.09616 [hep-th]}}.

\bibitem{Guica:2008mu}
M.~Guica, T.~Hartman, W.~Song, and A.~Strominger, ``{The Kerr/CFT Correspondence},'' \href{http://dx.doi.org/10.1103/PhysRevD.80.124008}{{\em Phys. Rev. D} {\bfseries 80} (2009) 124008}, \href{http://arxiv.org/abs/0809.4266}{{\ttfamily arXiv:0809.4266 [hep-th]}}.

\bibitem{Castro:2010fd}
A.~Castro, A.~Maloney, and A.~Strominger, ``{Hidden Conformal Symmetry of the Kerr Black Hole},'' \href{http://dx.doi.org/10.1103/PhysRevD.82.024008}{{\em Phys. Rev. D} {\bfseries 82} (2010) 024008}, \href{http://arxiv.org/abs/1004.0996}{{\ttfamily arXiv:1004.0996 [hep-th]}}.

\bibitem{Castro:2013lba}
A.~Castro, J.~M. Lapan, A.~Maloney, and M.~J. Rodriguez, ``{Black Hole Scattering from Monodromy},'' \href{http://dx.doi.org/10.1088/0264-9381/30/16/165005}{{\em Class. Quant. Grav.} {\bfseries 30} (2013) 165005}, \href{http://arxiv.org/abs/1304.3781}{{\ttfamily arXiv:1304.3781 [hep-th]}}.

\bibitem{Novaes:2014lha}
F.~Novaes and B.~Carneiro~da Cunha, ``{Isomonodromy, Painlevé Transcendents and Scattering off of Black Holes},'' \href{http://dx.doi.org/10.1007/JHEP07(2014)132}{{\em JHEP} {\bfseries 07} (2014) 132}, \href{http://arxiv.org/abs/1404.5188}{{\ttfamily arXiv:1404.5188 [hep-th]}}.

\bibitem{CarneirodaCunha:2015hzd}
B.~Carneiro~da Cunha and F.~Novaes, ``{Kerr Scattering Coefficients via Isomonodromy},'' \href{http://dx.doi.org/10.1007/JHEP11(2015)144}{{\em JHEP} {\bfseries 11} (2015) 144}, \href{http://arxiv.org/abs/1506.06588}{{\ttfamily arXiv:1506.06588 [hep-th]}}.

\bibitem{CarneirodaCunha:2015qln}
B.~Carneiro~da Cunha and F.~Novaes, ``{Kerr{\textendash}de Sitter greybody factors via isomonodromy},'' \href{http://dx.doi.org/10.1103/PhysRevD.93.024045}{{\em Phys. Rev. D} {\bfseries 93} no.~2, (2016) 024045}, \href{http://arxiv.org/abs/1508.04046}{{\ttfamily arXiv:1508.04046 [hep-th]}}.

\bibitem{Gamayun:2012ma}
O.~Gamayun, N.~Iorgov, and O.~Lisovyy, ``{Conformal field theory of Painlev{\'e} VI},'' \href{http://dx.doi.org/10.1007/JHEP10(2012)038}{{\em JHEP} {\bfseries 10} (2012) 038}, \href{http://arxiv.org/abs/1207.0787}{{\ttfamily arXiv:1207.0787 [hep-th]}}. [Erratum: JHEP 10, 183 (2012)].

\bibitem{Gamayun:2013auu}
O.~Gamayun, N.~Iorgov, and O.~Lisovyy, ``{How instanton combinatorics solves Painlev{\'e} VI, V and IIIs},'' \href{http://dx.doi.org/10.1088/1751-8113/46/33/335203}{{\em J. Phys. A} {\bfseries 46} (2013) 335203}, \href{http://arxiv.org/abs/1302.1832}{{\ttfamily arXiv:1302.1832 [hep-th]}}.

\bibitem{Litvinov:2013sxa}
A.~Litvinov, S.~Lukyanov, N.~Nekrasov, and A.~Zamolodchikov, ``{Classical Conformal Blocks and Painleve VI},'' \href{http://dx.doi.org/10.1007/JHEP07(2014)144}{{\em JHEP} {\bfseries 07} (2014) 144}, \href{http://arxiv.org/abs/1309.4700}{{\ttfamily arXiv:1309.4700 [hep-th]}}.

\bibitem{Iorgov:2014vla}
N.~Iorgov, O.~Lisovyy, and J.~Teschner, ``{Isomonodromic Tau-Functions from Liouville Conformal Blocks},'' \href{http://dx.doi.org/10.1007/s00220-014-2245-0}{{\em Commun. Math. Phys.} {\bfseries 336} no.~2, (2015) 671--694}, \href{http://arxiv.org/abs/1401.6104}{{\ttfamily arXiv:1401.6104 [hep-th]}}.

\bibitem{Aminov:2020yma}
G.~Aminov, A.~Grassi, and Y.~Hatsuda, ``{Black Hole Quasinormal Modes and Seiberg{\textendash}Witten Theory},'' \href{http://dx.doi.org/10.1007/s00023-021-01137-x}{{\em Annales Henri Poincare} {\bfseries 23} no.~6, (2022) 1951--1977}, \href{http://arxiv.org/abs/2006.06111}{{\ttfamily arXiv:2006.06111 [hep-th]}}.

\bibitem{Bonelli:2021uvf}
G.~Bonelli, C.~Iossa, D.~P. Lichtig, and A.~Tanzini, ``{Exact solution of Kerr black hole perturbations via CFT2 and instanton counting: Greybody factor, quasinormal modes, and Love numbers},'' \href{http://dx.doi.org/10.1103/PhysRevD.105.044047}{{\em Phys. Rev. D} {\bfseries 105} no.~4, (2022) 044047}, \href{http://arxiv.org/abs/2105.04483}{{\ttfamily arXiv:2105.04483 [hep-th]}}.

\bibitem{Bonelli:2022ten}
G.~Bonelli, C.~Iossa, D.~Panea~Lichtig, and A.~Tanzini, ``{Irregular Liouville Correlators and Connection Formulae for Heun Functions},'' \href{http://dx.doi.org/10.1007/s00220-022-04497-5}{{\em Commun. Math. Phys.} {\bfseries 397} no.~2, (2023) 635--727}, \href{http://arxiv.org/abs/2201.04491}{{\ttfamily arXiv:2201.04491 [hep-th]}}.

\bibitem{Lisovyy:2022flm}
O.~Lisovyy and A.~Naidiuk, ``{Perturbative connection formulas for Heun equations},'' \href{http://dx.doi.org/10.1088/1751-8121/ac9ba7}{{\em J. Phys. A} {\bfseries 55} no.~43, (2022) 434005}, \href{http://arxiv.org/abs/2208.01604}{{\ttfamily arXiv:2208.01604 [math-ph]}}.

\bibitem{fioravanti2021newmethodexactresults}
D.~Fioravanti and D.~Gregori, ``A new method for exact results on quasinormal modes of black holes,'' 2021.
\newblock \url{https://arxiv.org/abs/2112.11434}.

\bibitem{Kapec:2023ruw}
D.~Kapec, A.~Sheta, A.~Strominger, and C.~Toldo, ``{Logarithmic Corrections to Kerr Thermodynamics},'' \href{http://dx.doi.org/10.1103/PhysRevLett.133.021601}{{\em Phys. Rev. Lett.} {\bfseries 133} no.~2, (2024) 021601}, \href{http://arxiv.org/abs/2310.00848}{{\ttfamily arXiv:2310.00848 [hep-th]}}.

\bibitem{Kapec:2024zdj}
D.~Kapec, Y.~T.~A. Law, and C.~Toldo, ``{Quasinormal corrections to near-extremal black hole thermodynamics},'' \href{http://dx.doi.org/10.1007/JHEP06(2025)069}{{\em JHEP} {\bfseries 06} (2025) 069}, \href{http://arxiv.org/abs/2409.14928}{{\ttfamily arXiv:2409.14928 [hep-th]}}.

\bibitem{Rakic:2023vhv}
I.~Rakic, M.~Rangamani, and G.~J. Turiaci, ``{Thermodynamics of the near-extremal Kerr spacetime},'' \href{http://dx.doi.org/10.1007/JHEP06(2024)011}{{\em JHEP} {\bfseries 06} (2024) 011}, \href{http://arxiv.org/abs/2310.04532}{{\ttfamily arXiv:2310.04532 [hep-th]}}.

\bibitem{Kolanowski:2024}
M.~Kolanowski, D.~Marolf, I.~Rakic, M.~Rangamani, and G.~J. Turiaci, ``Looking at extremal black holes from very far away,'' \href{http://arxiv.org/abs/2409.16248}{{\ttfamily arXiv:2409.16248 [hep-th]}}.

\bibitem{Mariani:2025hee}
F.~Mariani and C.~Toldo, ``{Gravitational dynamics of near-extreme Kerr (Anti-)de Sitter black holes},'' \href{http://dx.doi.org/10.1007/JHEP02(2026)052}{{\em JHEP} {\bfseries 02} (2026) 052}, \href{http://arxiv.org/abs/2505.02674}{{\ttfamily arXiv:2505.02674 [hep-th]}}.

\bibitem{Castro:2025itb}
A.~Castro, R.~Mancilla, and I.~Papadimitriou, ``{Near-extremal dynamics away from the horizon},'' \href{http://dx.doi.org/10.1007/JHEP11(2025)083}{{\em JHEP} {\bfseries 11} (2025) 083}, \href{http://arxiv.org/abs/2507.01126}{{\ttfamily arXiv:2507.01126 [hep-th]}}.

\bibitem{PandoZayas:2026vbg}
L.~A. Pando~Zayas and J.~Zhang, ``{A Universality Theorem for the Quantum Thermodynamics of Near-Extremal Black Holes},'' \href{http://arxiv.org/abs/2602.16767}{{\ttfamily arXiv:2602.16767 [hep-th]}}.

\bibitem{Arnaudo:2025uos}
P.~Arnaudo, J.~Carballo, and B.~Withers, ``{Beyond quasinormal modes: a complete mode decomposition of black hole perturbations},'' \href{http://arxiv.org/abs/2510.18956}{{\ttfamily arXiv:2510.18956 [gr-qc]}}.

\bibitem{Arnaudo:2025kit}
P.~Arnaudo and B.~Withers, ``{Price's law from quasinormal modes},'' \href{http://arxiv.org/abs/2511.17703}{{\ttfamily arXiv:2511.17703 [gr-qc]}}.

\bibitem{Arnaudo:2026der}
P.~Arnaudo and B.~Withers, ``{Analytic structure of holographic thermal correlators from Fourier series},'' \href{http://arxiv.org/abs/2603.13469}{{\ttfamily arXiv:2603.13469 [hep-th]}}.

\bibitem{Aniceto:2026zmc}
I.~Aniceto, P.~Arnaudo, A.~Ratcliffe, and M.~Spali{\'n}ski, ``{Analytic approaches to perturbations of strongly coupled Yang-Mills plasma},'' \href{http://arxiv.org/abs/2606.12529}{{\ttfamily arXiv:2606.12529 [hep-th]}}.

\bibitem{Nekrasov:2011bc}
N.~Nekrasov, A.~Rosly, and S.~Shatashvili, ``{Darboux coordinates, Yang-Yang functional, and gauge theory},'' \href{http://dx.doi.org/10.1016/j.nuclphysbps.2011.04.150}{{\em Nucl. Phys. B Proc. Suppl.} {\bfseries 216} (2011) 69--93}, \href{http://arxiv.org/abs/1103.3919}{{\ttfamily arXiv:1103.3919 [hep-th]}}.

\bibitem{Denef:2009kn}
F.~Denef, S.~A. Hartnoll, and S.~Sachdev, ``{Black hole determinants and quasinormal modes},'' \href{http://dx.doi.org/10.1088/0264-9381/27/12/125001}{{\em Class. Quant. Grav.} {\bfseries 27} (2010) 125001}, \href{http://arxiv.org/abs/0908.2657}{{\ttfamily arXiv:0908.2657 [hep-th]}}.

\bibitem{Arnaudo:2024rhv}
P.~Arnaudo, G.~Bonelli, and A.~Tanzini, ``{One loop effective actions in Kerr-(A)dS black holes},'' \href{http://dx.doi.org/10.1103/PhysRevD.110.106006}{{\em Phys. Rev. D} {\bfseries 110} no.~10, (2024) 106006}, \href{http://arxiv.org/abs/2405.13830}{{\ttfamily arXiv:2405.13830 [hep-th]}}.

\bibitem{Arnaudo:2025btb}
P.~Arnaudo, G.~Bonelli, and A.~Tanzini, ``{One loop corrections to the thermodynamics of near-extremal Kerr-(A)dS black holes from Heun equation},'' \href{http://dx.doi.org/10.1007/JHEP12(2025)018}{{\em JHEP} {\bfseries 12} (2025) 018}, \href{http://arxiv.org/abs/2506.08959}{{\ttfamily arXiv:2506.08959 [hep-th]}}.

\bibitem{Teukolsky:1972my}
S.~A. Teukolsky, ``{Rotating black holes - separable wave equations for gravitational and electromagnetic perturbations},'' \href{http://dx.doi.org/10.1103/PhysRevLett.29.1114}{{\em Phys. Rev. Lett.} {\bfseries 29} (1972) 1114--1118}.

\bibitem{Batic:2007it}
D.~Batic and H.~Schmid, ``{Heun equation, Teukolsky equation, and type-D metrics},'' \href{http://dx.doi.org/10.1063/1.2720277}{{\em J. Math. Phys.} {\bfseries 48} (2007) 042502}, \href{http://arxiv.org/abs/gr-qc/0701064}{{\ttfamily arXiv:gr-qc/0701064}}.

\bibitem{Moitra:2019bub}
U.~Moitra, S.~K. Sake, S.~P. Trivedi, and V.~Vishal, ``{Jackiw-Teitelboim Gravity and Rotating Black Holes},'' \href{http://dx.doi.org/10.1007/JHEP11(2019)047}{{\em JHEP} {\bfseries 11} (2019) 047}, \href{http://arxiv.org/abs/1905.10378}{{\ttfamily arXiv:1905.10378 [hep-th]}}.

\bibitem{Arnaudo:2024bbd}
P.~Arnaudo, G.~Bonelli, and A.~Tanzini, ``{One-Loop Corrections to Near-Extremal Kerr Thermodynamics from Semiclassical Virasoro Blocks},'' \href{http://dx.doi.org/10.1103/cd6l-bl2s}{{\em Phys. Rev. Lett.} {\bfseries 134} no.~25, (2025) 251401}, \href{http://arxiv.org/abs/2412.16057}{{\ttfamily arXiv:2412.16057 [hep-th]}}.

\bibitem{Bac:2026eqj}
A.~Bac, A.~Castro, and D.~Jain, ``{Revisiting near-extremal and near-BPS black holes in AdS3 supergravity},'' \href{http://arxiv.org/abs/2604.24834}{{\ttfamily arXiv:2604.24834 [hep-th]}}.

\bibitem{Cheng:2026ull}
P.~Cheng and Y.-Q. Liu, ``{Universal Lichnerowicz Lifting of Near-Horizon Soft Modes},'' \href{http://arxiv.org/abs/2606.27308}{{\ttfamily arXiv:2606.27308 [hep-th]}}.

\bibitem{Despontin:2026xzg}
E.~Despontin, S.~Detournay, R.~Mancilla, and C.~Toldo, ``{Quantum corrections to the near-extremal thermodynamics of (warped) BTZ black holes},'' \href{http://arxiv.org/abs/2607.08482}{{\ttfamily arXiv:2607.08482 [hep-th]}}.

\bibitem{Bautista:2023sdf}
Y.~F. Bautista, G.~Bonelli, C.~Iossa, A.~Tanzini, and Z.~Zhou, ``{Black hole perturbation theory meets CFT2: Kerr-Compton amplitudes from Nekrasov-Shatashvili functions},'' \href{http://dx.doi.org/10.1103/PhysRevD.109.084071}{{\em Phys. Rev. D} {\bfseries 109} no.~8, (2024) 084071}, \href{http://arxiv.org/abs/2312.05965}{{\ttfamily arXiv:2312.05965 [hep-th]}}.

\bibitem{Cipriani:2025ikx}
A.~Cipriani, G.~Di~Russo, F.~Fucito, J.~F. Morales, H.~Poghosyan, and R.~Poghossian, ``{Resumming post-Minkowskian and post-Newtonian gravitational waveform expansions},'' \href{http://dx.doi.org/10.21468/SciPostPhys.19.2.057}{{\em SciPost Phys.} {\bfseries 19} no.~2, (2025) 057}, \href{http://arxiv.org/abs/2501.19257}{{\ttfamily arXiv:2501.19257 [gr-qc]}}.

\bibitem{CaronHuot:2025BornSeriesTidal}
S.~Caron-Huot, M.~Correia, G.~Isabella, and M.~Solon, ``Gravitational wave scattering via the born series: Scalar tidal matching to $o(g^7)$ and beyond,'' \href{http://arxiv.org/abs/2503.13593}{{\ttfamily arXiv:2503.13593 [hep-th]}}.

\bibitem{CombaluzierSzteinsznaider:2026DynamicalTidal}
O.~Combaluzier-Szteinsznaider, D.~Glazer, A.~Joyce, M.~J. Rodriguez, and L.~Santoni, ``Dynamical tidal response of schwarzschild black holes,'' \href{http://arxiv.org/abs/2511.02372}{{\ttfamily arXiv:2511.02372 [gr-qc]}}.

\bibitem{Lestingi:2026peq}
J.~Lestingi, L.~Sberna, and S.~R. Green, ``{Schr{\"o}dinger perturbation theory for black hole quasinormal modes},'' \href{http://arxiv.org/abs/2607.19492}{{\ttfamily arXiv:2607.19492 [gr-qc]}}.

\bibitem{Matone:1995rx}
M.~Matone, ``{Instantons and recursion relations in N=2 SUSY gauge theory},'' \href{http://dx.doi.org/10.1016/0370-2693(95)00920-G}{{\em Phys. Lett. B} {\bfseries 357} (1995) 342--348}, \href{http://arxiv.org/abs/hep-th/9506102}{{\ttfamily arXiv:hep-th/9506102}}.

\bibitem{Arnaudo:2024sen}
P.~Arnaudo and B.~Withers, ``{Exact low-temperature Green{\textquoteright}s functions in AdS/CFT: From the Heun equation to the confluent Heun equation},'' \href{http://dx.doi.org/10.1103/8n3f-2d33}{{\em Phys. Rev. D} {\bfseries 111} no.~12, (2025) L121903}, \href{http://arxiv.org/abs/2412.01923}{{\ttfamily arXiv:2412.01923 [hep-th]}}.

\bibitem{Berti:2009kk}
E.~Berti, V.~Cardoso, and A.~O. Starinets, ``{Quasinormal modes of black holes and black branes},'' \href{http://dx.doi.org/10.1088/0264-9381/26/16/163001}{{\em Class. Quant. Grav.} {\bfseries 26} (2009) 163001}, \href{http://arxiv.org/abs/0905.2975}{{\ttfamily arXiv:0905.2975 [gr-qc]}}.

\bibitem{suzuki:1998vy}
H.~Suzuki, E.~Takasugi, and H.~Umetsu, ``{Perturbations of Kerr-de Sitter black hole and Heun's equations},'' \href{http://dx.doi.org/10.1143/PTP.100.491}{{\em Prog. Theor. Phys.} {\bfseries 100} (1998) 491--505}, \href{http://arxiv.org/abs/gr-qc/9805064}{{\ttfamily arXiv:gr-qc/9805064}}.

\bibitem{Novaes:2018fry}
F.~Novaes, C.~Marinho, M.~Lencs{\'e}s, and M.~Casals, ``{Kerr-de Sitter Quasinormal Modes via Accessory Parameter Expansion},'' \href{http://dx.doi.org/10.1007/JHEP05(2019)033}{{\em JHEP} {\bfseries 05} (2019) 033}, \href{http://arxiv.org/abs/1811.11912}{{\ttfamily arXiv:1811.11912 [gr-qc]}}.

\bibitem{Denef:2000nb}
F.~Denef, ``{Supergravity flows and D-brane stability},'' \href{http://dx.doi.org/10.1088/1126-6708/2000/08/050}{{\em JHEP} {\bfseries 08} (2000) 050}, \href{http://arxiv.org/abs/hep-th/0005049}{{\ttfamily arXiv:hep-th/0005049}}.

\bibitem{Denef:2002ru}
F.~Denef, ``{Quantum quivers and Hall / hole halos},'' \href{http://dx.doi.org/10.1088/1126-6708/2002/10/023}{{\em JHEP} {\bfseries 10} (2002) 023}, \href{http://arxiv.org/abs/hep-th/0206072}{{\ttfamily arXiv:hep-th/0206072}}.

\bibitem{Bates:2003vx}
B.~Bates and F.~Denef, ``{Exact solutions for supersymmetric stationary black hole composites},'' \href{http://dx.doi.org/10.1007/JHEP11(2011)127}{{\em JHEP} {\bfseries 11} (2011) 127}, \href{http://arxiv.org/abs/hep-th/0304094}{{\ttfamily arXiv:hep-th/0304094}}.

\bibitem{Denef:2007vg}
F.~Denef and G.~W. Moore, ``{Split states, entropy enigmas, holes and halos},'' \href{http://dx.doi.org/10.1007/JHEP11(2011)129}{{\em JHEP} {\bfseries 11} (2011) 129}, \href{http://arxiv.org/abs/hep-th/0702146}{{\ttfamily arXiv:hep-th/0702146}}.

\bibitem{Arnaudo:2025kof}
P.~Arnaudo, A.~Grassi, and Q.~Hao, ``{On quivers, spectral networks and black holes},'' \href{http://arxiv.org/abs/2502.01526}{{\ttfamily arXiv:2502.01526 [hep-th]}}.

\bibitem{Ivo:2025yek}
V.~Ivo, J.~Maldacena, and Z.~Sun, ``{Physical instabilities and the phase of the Euclidean path integral},'' \href{http://dx.doi.org/10.1007/JHEP04(2026)118}{{\em JHEP} {\bfseries 04} (2026) 118}, \href{http://arxiv.org/abs/2504.00920}{{\ttfamily arXiv:2504.00920 [hep-th]}}.

\bibitem{Turiaci:2025xwi}
G.~J. Turiaci and C.-H. Wu, ``{The wavefunction of a quantum S$^{1}$ {\texttimes} S$^{2}$ universe},'' \href{http://dx.doi.org/10.1007/JHEP07(2025)158}{{\em JHEP} {\bfseries 07} (2025) 158}, \href{http://arxiv.org/abs/2503.14639}{{\ttfamily arXiv:2503.14639 [hep-th]}}.

\bibitem{Alday:2009aq}
L.~F. Alday, D.~Gaiotto, and Y.~Tachikawa, ``{Liouville Correlation Functions from Four-dimensional Gauge Theories},'' \href{http://dx.doi.org/10.1007/s11005-010-0369-5}{{\em Lett. Math. Phys.} {\bfseries 91} (2010) 167--197}, \href{http://arxiv.org/abs/0906.3219}{{\ttfamily arXiv:0906.3219 [hep-th]}}.

\bibitem{Alday:2009fs}
L.~F. Alday, D.~Gaiotto, S.~Gukov, Y.~Tachikawa, and H.~Verlinde, ``{Loop and surface operators in N=2 gauge theory and Liouville modular geometry},'' \href{http://dx.doi.org/10.1007/JHEP01(2010)113}{{\em JHEP} {\bfseries 01} (2010) 113}, \href{http://arxiv.org/abs/0909.0945}{{\ttfamily arXiv:0909.0945 [hep-th]}}.

\end{thebibliography}

\begingroup\raggedright\endgroup

\end{document}